\documentstyle[aps,prd,eqsecnum,amssymb,amsfonts,floats,epsfig]{revtex}
\newcommand{\abar}{{\overline{\alpha}}}
\newcommand{\abs}[1]{\left|#1\right|}
\newcommand{\trans}[1]{{#1}^{\text{tr}}}
\newcommand{\eqref}[1]{(\ref{#1})}
\newcommand{\del}{\bbox{\nabla}}
\newcommand{\rvert}{|}

\newcommand{\lbp}{\bbox{(}}
\newcommand{\rbp}{\bbox{)}}

\newcommand{\kay}{{{\bf k}}}

\newcommand{\q}{{{\bf q}}}
\newcommand{\x}{{{\bf x}}}

\newcommand{\D}{{\cal D}}
\newcommand{\Dend}{{d}}
\newcommand{\mc}[1]{{\cal #1}}
\newcommand{\wt}[1]{\widetilde{#1}}

\newcommand{\R}{{\Bbb R}}
\newcommand{\C}{{\Bbb C}}

\newcommand{\Ddim}{D}
\newcommand{\M}{\kay}
\newcommand{\N}{\q}
\newcommand{\off}{{0+\phi\phi}}
\newcommand{\fff}{{\phi\phi\phi}}

\newcommand{\Mex}{M_{\text{ex}}}
\newcommand{\Oo}{\Omega_0}
\newcommand{\Tr}{\mathop{\rm Tr}\nolimits}
\newcommand{\Real}{\mathop{\rm Re}\nolimits}
\newcommand{\Imag}{\mathop{\rm Im}\nolimits}
\newcommand{\diag}{\mathop{\rm diag}\nolimits}
\newcommand{\coord}{co\-or\-di\-nate}

\newcommand{\Arune}{{\frak A}}
\newcommand{\Brune}{{\frak B}}
\newcommand{\Crune}{{\frak C}}
\newcommand{\tag}[1]{\eqnum{\protect{#1}}}
\newcommand{\iint}{\int\!\!\!\!\int}
\newcommand{\iOint}{\int\limits_{\,\Omega}\!\!\!\!\int}
\newcommand{\tbinom}[2]{{\textstyle #1\choose #2}}

\begin{document}
\title{Modeling the Decoherence of Spacetime\thanks{\textbf{Copyright
      The American Physical Society 1997.  All rights reserved.
      Except as provided under U.S.\ copyright law, this work may not
      be reproduced, resold, distributed or modified without the
      express permission of The American Physical Society.  The
      archival version of this work was published in \textit{Phys.\ 
        Rev.\ D} \underline{57}, 768 (1998).}}}  
\author{John T.~Whelan\thanks{Electronic address: whelan@physics.utah.edu}}
\address{Department of Physics, University of Utah,
  115~S.~1400~E.~Room 201, Salt Lake City, UT 84112-0830}
\date{Received 4 June 1997; published 22 December 1997} \maketitle
\begin{abstract}
The question of whether unobserved short-wavelength modes of the
gravitational field can induce decoherence in the
long-wavelength modes (``the decoherence of spacetime'') is
addressed using a simplified model of perturbative general relativity,
related to the Nordstr\"{o}m-Einstein-Fokker theory, where the metric
is assumed to be conformally flat.
 For some long-wavelength
coarse grainings, the Feynman-Vernon influence phase is found to be
effective at suppressing the off-diagonal elements of the decoherence
functional.  The requirement that the short-wavelength modes be in a
sufficiently high-temperature state places limits on the applicability
of this perturbative approach.
\end{abstract}


\section{Introduction}

        Any theory of quantum cosmology, which treats the entire
universe as a quantum-mechanical system, should predict classical
behavior in the regimes where we know classical physics to be valid.
In particular, a quantum theory of gravity should predict classical
spacetime on macroscopic scales.  One way of formulating the quantum
mechanics of a closed system ({e.g.,} the universe) is
generalized quantum mechanics \cite{gqm}, in which probabilities are
assigned to alternatives (outcomes of a series of observations) only
if the quantum-mechanical interference between pairs of alternatives
vanishes.  This non-interference, known as decoherence, is a minimum 
condition for classical behavior.

        As described in Sec.~\ref{sec:infl}, The physical process
associated with decoherence \cite{zurek} occurs by interaction of the
system of interest with an environment about which no information is
gained.  It has been conjectured that long-wavelength features of the
gravitational field may be made to decohere by their interactions with
the short-wavelength modes of the field, thus allowing classical
behavior of the gravitational field when observed on large scales.

        This paper provides evidence for this phenomenon, the
decoherence of spacetime.  This differs from previous work
\cite{JJH89,padman} which used an additional field to obtain
decoherence of the gravitational field in cosmological models, in that
the decoherence examined here is induced in a field theory
representing only gravity with no external matter field.

        The theory considered in this work is a scalar field theory
with a self-interaction similar to that of the metric in a
perturbative expansion of the action for General Relativity (GR).
 As described in Sec.~\ref{sec:GRscalar}, it is also the perturbative
form of the Nordstr\"{o}m-Einstein-Fokker theory \cite{NEF} of a
conformally flat metric, if the scalar field is defined proportional
to the deviation from unity of some power of the conformal factor,
such as the scaling of the volume element.

        Section~\ref{sec:div} demonstrates the effects of splitting
the scalar field into long-wavelength and short-wavelength parts, and
classifies the terms in the action by the number of short-wavelength
modes (SWMs).  The trace over the SWMs is complicated by the presence
of a cubic term in the action.

        Temporarily removing the terms with one and three SWM factors
leaves an action whose terms are all quadratic in the SWMs, or
independent of them.  Thus the trace over the SWMs can be performed
explicitly, and this is done in Sec.~\ref{sec:quad}.

        As described in Sec.~\ref{ssec:breakdown}, the perturbative
corrections to the decoherence functional can cause elements (namely
those representing quantum interference) which are finite in the
non-interacting theory to be suppressed if the SWMs are in a thermal
state whose temperature is sufficiently high.  Then certain terms in
the perturbation series can become large in the high-temperature
limit, producing seemingly non-perturbative effects, including
decoherence.

        Section~\ref{sec:full} summarizes the results of
Appendix~\ref{app:full}, that reinserting the terms with one and
three SWM factors into the action has no substantial effect on the
result of Sec.~\ref{sec:quad}.  The terms linear in the SWMs can be
removed by completion of the square to recover the original result.
The terms cubic in the SWMs are examined in a perturbation series, and
each term is seen to be perturbatively finite, even in the
high-temperature limit.  So, according to the perturbative analysis,
the effect of the cubic terms is to multiply the influence functional
by a factor of order unity.

        Section~\ref{sec:interp} applies the properties of the
decoherence functional found in Sec.~\ref{sec:quad} to a class of
practical coarse grainings.  First, in Sec.~\ref{ssec:modes}, we show
that the suppression factor enforcing decoherence involves Fourier
components of the field whose temporal frequency is less than their
spatial frequency.  In Sec.~\ref{ssec:CG}, we construct the coarse
grainings of interest.  Taking the temperature of the short-wavelength
background to be that of the cosmic graviton background requires that
we make the division into ``short'' and ``long'' wavelengths at or
above the millimeter scale in order to use the ``high temperature''
approximation described in Sec.~\ref{ssec:breakdown}.  Constructing a
coarse graining by a weighted field average which describes a group of
modes in the appropriate region, we obtain decoherence if the
spacetime region is large enough, and the environment temperature is
high enough, compared with the scale set by the oscillation of the
weighting function.

        Finally, Sec.~\ref{ssec:imprac} shows that the essential
feature of this decoherence mechanism is the system-environment split,
and not the long-wavelength nature of the system, by showing that
coarse grainings referring only to \emph{short}-wavelength features
can be made to decohere by their interaction with the
\emph{long}-wavelength ``environment''.

        For reference, a summary of the important notation used in the
body of the paper is provided in Appendix~\ref{app:gloss}.

\section{Environment-induced decoherence
and the influence phase}\label{sec:infl}

\subsection{Generalized Quantum Mechanics}

        In a sum-over-histories quantum mechanics, the natural
definition of the probability $p(\alpha)$ that a certain alternative
$c_\alpha$ is realized is as the square of an amplitude, which is
constructed via a sum of $e^{i\text{ action}}$ only over those
histories which are in the class corresponding to that alternative.
However, probabilities defined in this way will in general not obey
the probability sum rule, {i.e.,} the probability calculated
for a class which is the union of two disjoint smaller classes
($c_{\overline{\alpha}}=c_\alpha\cup c_{\alpha'}$, $c_\alpha\cap
c_{\alpha'}=\emptyset$) will in general not be the sum of the
probabilities for those two classes [$p(\overline{\alpha})\ne
p(\alpha)+p(\alpha')$].  Generalized quantum mechanics (GQM) addresses
this problem by replacing the probability $p(\alpha)$ of a single
alternative with a decoherence functional $D(\alpha,\alpha')$ defined
on each pair of alternatives.  When an exhaustive set of mutually
exclusive classes $\{c_\alpha\}$ has the property that
$D(\alpha,\alpha')=0$ for $\alpha\ne\alpha'$, known as decoherence,
one can then identify the diagonal elements of the decoherence
functional (thought of as a matrix) as the probabilities
$p(\alpha)=D(\alpha,\alpha)$ for the alternatives in that set.

        A sum-over-histories generalized quantum mechanics, as
formulated by Hartle \cite{gqm}, requires three elements:

         (1) A definition of the \emph{fine-grained histories},
$\{h\}$, the most precise descriptions of the state of the system.
For example, these may be particle paths $q(t)$ or field histories
$\varphi(x)$ over spacetime, 

        (2) A rule for partitioning those fine-grained histories
into \emph{coarse-grained} classes or alternatives $\{c_\alpha\}$, and

        (3) A \emph{decoherence functional} $D[h,h']$ defined
on pairs of histories (fine- or coarse-grained).

The decoherence functional must obey the following four conditions:

\begin{mathletters}\label{dfrules}
      ``Hermiticity'':\label{herm}
          \begin{equation}
          D(\alpha',\alpha)=D(\alpha,\alpha')^*;
          \end{equation}

      positivity of diagonal elements:
          \begin{equation}\label{pos}
          D(\alpha,\alpha)\ge 0;
          \end{equation}

      normalization:
          \begin{equation}
          \sum_{\alpha}\sum_{\alpha'} D(\alpha,\alpha')=1;
          \end{equation}

      superposition: 
If $\{c_\abar\}$ is a coarse graining constructed by
combining classes in $\{c_\alpha\}$ to form larger classes (``coarser
graining''), i.e., $c_\abar =
\bigcup\limits_{\alpha\in\abar} c_\alpha$, the decoherence
functional for $\{c_\abar\}$ can be constructed from the
one for $\{c_\alpha\}$ by
\begin{equation}\label{super}
D(\abar,{\abar'})
=\sum_{\alpha\in\abar}
\sum_{\alpha'\in\abar'}
D(\alpha,\alpha').
\end{equation}
\end{mathletters}
Note that the superposition law \eqref{super} allows one to construct
the coarse-grained decoherence functional from the fine-grained one,
via
\begin{equation}
D(\alpha,\alpha')
=\sum_{h\in\alpha}\sum_{h'\in\alpha'}D[h,h'].
\end{equation}

        When the initial state is described by a normalized density
matrix $\rho$, and there is no specified final state, the fine-grained
decoherence functional for a field $\varphi$ with action $S$ is given
by\footnote{Throughout this paper, we will use units in which
$\hbar=1=c$.}
\begin{equation}
D[\varphi,\varphi']
=\rho(\varphi_i,\varphi'_i)\,\delta(\varphi'_f-\varphi_f)
e^{i(S[\varphi]-S[\varphi'])}.
\end{equation}

\subsection{The Influence Phase}

        Decoherence in most physical systems is caused by a division
into the ``system'' of interest, and an ``environment'' about which no
information is gathered.  In the language of generalized quantum
mechanics, this means that the coarse graining is described by
alternatives which refer only to the system variables.  (See
\cite{classeq} for further details and a bibliography of prior work.)

        If we make a division of $\varphi$ into system variables
$\Phi$ and environment variables $\phi$, split up the action into a
$\phi$-independent piece $S_\Phi[\Phi]=S\rvert_{\phi=0}$ and a piece
$S_E$ describing the environment and its interaction with the system:
\begin{equation}
S[\varphi]=S_\Phi[\Phi]+S_E[\phi,\Phi],
\end{equation}
and assume that the initial state is the product of uncorrelated
states for the system and the environment:
\begin{equation}\label{rhosep}
\rho(\varphi_i,\varphi'_i)
=\rho_\Phi(\Phi_i,\Phi'_i)\rho_\phi(\phi_i,\phi'_i),
\end{equation}
then the decoherence functional for a coarse-graining which makes no
reference to the environment variables (but is still fine-grained in
the system variables) can be written
\begin{mathletters}\label{decinfl}
\begin{equation}
D[\Phi,\Phi']=\int\D\phi\,\D\phi' D[\varphi,\varphi']
=\rho_\Phi(\Phi_i,\Phi'_i)\delta(\Phi'_f-\Phi_f)
e^{i(S_\Phi[\Phi]-S_\Phi[\Phi']+W[\Phi,\Phi'])}
\end{equation}
where
\begin{equation}\label{infl}
e^{iW[\Phi,\Phi']}
=\int\D\phi\,\D\phi' 
\rho_\phi(\phi_i,\phi'_i)\delta(\phi'_f-\phi_f)
e^{i(S_E[\phi,\Phi]-S_E[\phi',\Phi'])}.
\end{equation}
\end{mathletters}
$W[\Phi,\Phi']$ is called the Feynman-Vernon influence
phase \cite{feynvern}; if the \emph{influence functional} $e^{iW}$
becomes small for $\Phi\neq\Phi'$, the ``off-diagonal'' parts of
$D[\Phi,\Phi']$ will be suppressed, causing alternatives defined in
terms of $\Phi$ to decohere \cite{classeq}.

\section{A scalar field theory modeling the
gravitational interaction}\label{sec:GRscalar}

        The goal of this work is to perform the division described in
Sec.~\ref{sec:infl} on a theory modeling vacuum gravity, with the
short-wavelength modes acting as the environment which induces
decoherence in the long-wavelength system.  The idea behind this is
that for coarse grainings which deal only with averages over
sufficiently large regions of spacetime, gravity should behave
classically, and thus such coarse grainings should decohere.

        In order to model the self-interaction of the gravitational
field without dealing with the gauge-fixing and other complications
arising from the tensor nature of the metric in General Relativity, we
will consider a toy model of a single scalar field $\varphi$ moving on
$D+1$-dimensional Minkowski space\footnote{As discussed in
Sec.~\ref{sssec:scale}, this should be a reasonable assumption in a
cosmological scenario if the length scales in the problem do not
approach the Hubble scale $cH_0^{-1}$.  Note also that an analogous
scalar field model, with a different Ricci-flat background metric
$q_{ab}$ in place of the Minkowski metric $\eta_{ab}$, can also be
constructed.  \emph{That} action is obtained by a perturbative
expansion of the Einstein-Hilbert action when the metric is required
to be conformally related to the background: $g_{ab}=\Omega^2
q_{ab}$.}  with the action
\begin{equation}\label{toyact}
S[\varphi]=-\frac{1}{2}\int d^{\Ddim+1}\!x\,
[1-(2\pi)^{\Ddim/2}\ell\varphi](\nabla\varphi)^2.
\end{equation}

        This scalar field theory is a promising toy model for
perturbative general relativity for two reasons.  First, 
consider the Einstein-Hilbert action
\begin{equation}\label{EHact}
S=-\frac{1}{16\pi G}\int d^{\Ddim+1}\!x\sqrt{\abs{g}}\,R
\end{equation}
and define the difference $\gamma_{ab}=g_{ab}-\eta_{ab}$ between the
actual metric and a flat background metric.  If we perform a
perturbative expansion\footnote{Because quantities like the inverse
metric $g^{ab}$ cannot be expressed in closed form in terms of
$\gamma_{ab}$, the action will become an infinite series of terms
containing increasing powers of $\gamma_{ab}$.  One can treat the
theory perturbatively, but it is not clear that $\gamma_{ab}$ is the
most physically relevant quantity in which to carry out that
expansion.} \cite{dewitt} of \eqref{EHact} in powers of $\gamma_{ab}$,
the lowest order terms have two powers of $\gamma_{ab}$ and two
derivatives and describe free wave propagation, while the first
self-interaction terms have three powers of $\gamma_{ab}$ and two
derivatives.  If we replace the tensor-valued $\gamma_{ab}$ by a
scalar field $\varphi$, the most general action which has this form is
\eqref{toyact}.

        Second, \eqref{toyact} can also be obtained by perturbative
expansion of the action for the Nordstr\"{o}m-Einstein-Fokker theory
\cite{NEF} of gravity, which is given by the Einstein-Hilbert action
\eqref{EHact} restricted to conformally flat metrics
\begin{equation}\label{confmet}
g_{ab}=\Omega^2 \eta_{ab}.
\end{equation}
The classical equation for $\Omega$ is obtained by varying the action
\begin{equation}
S[\Omega]
=-\frac{4\Ddim}{\Ddim-1}\int d^{\Ddim+1}\!x
\left(\nabla\Omega^{\frac{\Ddim-1}{2}}\right)^2.\label{omact}
\end{equation}
In $3+1$ spacetime dimensions ($D=3$), this corresponds to the
statement that the conformal factor $\Omega$ behaves as a free
massless scalar field.

        However, it may be that a quantity defining a useful coarse
graining is proportional to some power $\Omega^\nu$ of the
conformal factor.  For example, the volume of a spacetime region $S$
in the metric \eqref{confmet} will be
\begin{equation}\label{vol}
V=\int\limits_S d^{\Ddim+1}\!x\sqrt{\abs{g}}
=\int\limits_S d^{\Ddim+1}\!x\,\Omega^{D+1},
\end{equation}
since the metric determinant in this theory is given by
$-(\Omega^2)^{D+1}$.  It is thus useful to express the theory
in terms of\footnote{The fact that $\zeta$ runs from $-1$ to $\infty$
(for positive $\nu$) as $\Omega$ runs from $0$ to $\infty$ should not
be cause for alarm, as the choice of $\zeta$ rather than $1+\zeta$ is
tailored to expanding the action for small $\zeta$.}
\begin{equation}\label{zetadef}
\zeta=\Omega^\nu-1
\end{equation}
so that $\zeta=0$ corresponds to no deviation from the flat background
metric.

  Then we obtain a
self-interacting\footnote{\protect{\label{fn:free}}It is a common
phenomenon that a theory which is free in one set of variables may
exhibit interactions and the resulting decoherence when described in
terms of another set.  The theories considered in \cite{brand,feld}
are of this sort, as are the linear oscillator models of \cite{C&L}
when expressed in terms of normal modes.}  action, which when expanded
perturbatively for $\zeta\ll 1$ becomes
\begin{equation}
S[\zeta]
\approx-\frac{\Ddim(\Ddim-1)}{\nu^2}
\int d^{\Ddim+1}\!x
\left(1-\frac{2\nu+1-\Ddim}{\nu}\zeta\right)(\nabla\zeta)^2.
\end{equation}
Rescaling to define
\begin{equation}
\varphi=\frac{\zeta}{2\nu\ell_p}\sqrt{\frac{\Ddim(\Ddim-1)}{2\pi}}
\end{equation}
and
\begin{equation}
\ell=\ell_p\frac{2(2\nu+1-\Ddim)}{\sqrt{\Ddim(\Ddim-1)(2\pi)^{\Ddim-1}}},
\end{equation}
where $\ell_p=G^{1/2}$ is the Planck length, we obtain the action
\eqref{toyact} for the scalar field $\varphi$.
Note that if
$\nu=(\Ddim-1)/2$, then the coupling constant $\ell$ vanishes, and we
have a free theory as dictated by \eqref{omact}.

        If we take $\nu=D+1$ in our definition \eqref{zetadef} of
$\zeta$, the volume \eqref{vol} of our spacetime region, when offset
and rescaled by a reference volume $V_0$ (which we take to be the
volume $\int_S d^{\Ddim+1}\!x$ of the region in the
background metric) will be
\begin{equation}
\frac{V-V_0}{V_0}
=\frac{\int\limits_S d^{\Ddim+1}\!x
\left(\sqrt{\abs{g}}-1\right)}
{\int\limits_S d^{\Ddim+1}\!x}
=\frac{\int\limits_S d^{\Ddim+1}\!x\,\zeta}
{\int\limits_S d^{\Ddim+1}\!x}
=\langle{\zeta}\rangle_S
=\frac{(\Ddim+1)(2\pi)^{\Ddim/2}}{\Ddim+3}\langle\ell\varphi\rangle_S,
\label{volnorm}\end{equation}
a field average of $\varphi$ over the region $S$.

        Note that the quantization of the action \eqref{toyact},
obtained from the Einstein-Hilbert action restricted to conformally
flat metrics, is \emph{not} the same as quantization of the conformal
modes of general relativity, since we have made the restriction
\eqref{confmet} at the level of the action, producing a theory which
has no other degrees of freedom.  This scalar theory of gravity is,
however, a geometric theory similar to GR, with a similar
self-interaction, but without many of the complications of the full
theory.

\section{Dividing the modes}\label{sec:div}

        We want to make a division of the field $\varphi$
appearing in the action
\begin{mathletters}
\begin{eqnarray}
S[\varphi]&=&\int_{-T/2}^{T/2}dt\, L(t)
\\
L(t)&=&\frac{1}{2}\int d^\Ddim\!x\, [1-(2\pi)^{\Ddim/2}\ell\varphi]
[\dot{\varphi}^2-(\del\varphi)^2]
\end{eqnarray}
\end{mathletters}
into long-wavelength modes (LWMs), labelled by $\Phi$, to act as the
``system'' and short-wavelength modes (SWMs), labelled by $\phi$, to
act as the ``environment''.  For reasons of mathematical convenience,
we first make this division only in the spatial directions.  First we
reexpress the Lagrangian in terms of the Fourier transform
\begin{mathletters}\label{fourier}
\begin{eqnarray}
\varphi_\kay(t)&=&\int\frac{d^\Ddim\!x}{(2\pi)^{\Ddim/2}}
e^{-i\kay\cdot\x}\varphi({\bf x},t);
\\
\varphi({\bf x},t)&=&\int\frac{d^\Ddim\!k}{(2\pi)^{\Ddim/2}}
e^{i\x\cdot\kay}\varphi_\kay(t)
\end{eqnarray}
\end{mathletters}
to get
\begin{equation}
L(t)=\frac{1}{2}\int d^\Ddim\!k\, 
(\abs{\dot{\varphi}}^2-\kay^2\abs{\varphi}^2)
-\frac{\ell}{2}\int
d^\Ddim\!k_1\, d^\Ddim\!k_2\, d^\Ddim\!k_3\,
\delta^\Ddim\!(\kay_1+\kay_2+\kay_3)
(\varphi_{1}\dot{\varphi}_{2}\dot{\varphi}_{3}
+\kay_2\cdot\kay_3\varphi_{1}\varphi_{2}\varphi_{3}),
\end{equation}
where we have streamlined the notation by writing
$\varphi=\varphi_\kay$, $\varphi_1=\varphi_{\kay_1}$, etc.

        We define the long-wavelength sector $\mc{L}=\{\N\,|\,q<k_c\}$
and the short-wavelength sector $\mc{S}=\{\M\,|\,k>k_c\}$, and define
the long- and short-wavelength modes by
\begin{mathletters}\label{Phiphi}
\begin{eqnarray}
\Phi_\N(t)&=&\varphi_\N(t)\Theta(k_c-q)\qquad\textbf{LWM} \\
\phi_\M(t)&=&\varphi_\M(t)\Theta(k-k_c)\qquad\textbf{SWM} 
\end{eqnarray}
\end{mathletters}
so that the part of the Lagrangian quadratic in $\varphi$ becomes
\begin{equation}
\frac{1}{2}\int d^\Ddim\!x\,[\dot{\varphi}^2-(\del\varphi)^2]
=\frac{1}{2}\int\limits_{\mc{L}} d^\Ddim\!q\,
\left(
        \abs{\dot{\Phi}}^2-\N^2\abs{\Phi}^2
\right)
+\frac{1}{2}\int\limits_{\mc{S}} d^\Ddim\!k\,
\left(
        \abs{\dot{\phi}}^2-\M^2\abs{\phi}^2
\right).
\end{equation}
Taking into account the fact that $\varphi(\x)$ is real, which means
$\varphi_{-\kay}=\varphi_\kay^*$, or $\Phi_{-\N}=\Phi_\N^*$ and
$\phi_{-\M}=\phi_\M^*$, we can write any expression using only half of
the complex modes, which define the other half by complex conjugation.
We define $\mc{L}/2$ and $\mc{S}/2$ as arbitrarily chosen
halves of $\mc{L}$ and $\mc{S}$ so that
$\{\Phi_\N|\N\in\mc{L}/2\}$ and $\{\phi_\M|\M\in\mc{S}/2\}$
between them define $\varphi_\kay$.  This makes the noninteracting
($\ell=0$) action
\begin{eqnarray}
\frac{1}{2}\int d^\Ddim\!x\,[\dot{\varphi}^2-(\del\varphi)^2]
=\int\limits_{\mc{L}/2} d^\Ddim\!q\,
\left(
        \abs{\dot{\Phi}_\N}^2-\q^2\abs{{\Phi}_\N}^2
\right)
+\int\limits_{\mc{S}/2} d^\Ddim\!k\,
\left(
        \abs{\dot{\phi}_\M}^2-\kay^2\abs{{\phi}_\M}^2
\right).
\end{eqnarray}
which is the action of a set of uncoupled harmonic oscillators.  The
interaction terms can be classified by the number of factors of the
``environment'' field $\phi$ to give
\begin{equation}\label{Ldiv}
L[\varphi]=L[\phi,\Phi]
=L_\Phi[\Phi]+L_0[\phi]+\ell L_\phi[\phi,\Phi]
+\ell L_{\phi\phi}[\phi,\Phi]+\ell L_\fff[\phi],
\end{equation}
where 
\begin{mathletters}\label{Lparts}
\begin{eqnarray}
L_\Phi[\Phi]&=&\int\limits_{\mc{L}/2} d^\Ddim\!q\,
\left(
        \abs{\dot{\Phi}}^2-\q^2\abs{\Phi}^2
\right)
-\frac{\ell}{2}\int d^\Ddim\!q_1\, d^\Ddim\!q_2\, d^\Ddim\!q_3\,
\delta^\Ddim\!(\N_1+\N_2+\N_3)
\left(
        \Phi_{1}\dot{\Phi}_{2}\dot{\Phi}_{3}
        +\N_{2}\cdot\N_{3}\Phi_{1}\Phi_{2}\Phi_{3}
\right)
\\
\label{L0}
L_0[\phi]&=&\int\limits_{\mc{S}/2} d^\Ddim\!k\,
\left(
        \abs{\dot{\phi}}^2-\M^2\abs{\phi}^2
\right)
\\
\label{Lphi}
 L_{\phi}[\phi,\Phi]&=&
 -\frac{1}{2}\int d^\Ddim\!k\, d^\Ddim\!q_1\, d^\Ddim\!q_2\,
 \delta^\Ddim\!(\M+\N_1+\N_2)
\left[
        \phi\dot{\Phi}_{1}\dot{\Phi}_{2}
         +2\dot{\phi}\dot{\Phi}_{1}\Phi_{2}
         +(2\M+\N_{1})\cdot\N_{2}\phi\Phi_{1}\Phi_{2}
\right]
\\
\label{Lphiphi}
 L_{\phi\phi}[\phi,\Phi]&=&
 -\frac{1}{2}\int d^\Ddim\!q\, d^\Ddim\!k_1\, d^\Ddim\!k_2\,
 \delta^\Ddim\!(\M_1+\M_2+\N)
\left[
        \Phi\dot{\phi}_{1}\dot{\phi}_{2}
         +2\dot{\Phi}\dot{\phi}_{1}\phi_{2}
         +(2\N+\M_{1})\cdot\M_{2}\Phi\phi_{1}\phi_{2}
\right]
\\
\label{Lphiphiphi}
L_\fff[\phi]&=&
-\frac{1}{2}\int d^\Ddim\!k_1\, d^\Ddim\!k_2\, d^\Ddim\!k_3\,
\delta^\Ddim\!(\M_1+\M_2+\M_3)
\left(
        \phi_{1}\dot{\phi}_{2}\dot{\phi}_{3}
        +\M_{2}\cdot\M_{3}\phi_{1}\phi_{2}\phi_{3}
\right).
\end{eqnarray}
\end{mathletters}
Under this division of the field $\varphi$, we can perform the
division of the decoherence functional described by \eqref{decinfl},
with
\begin{equation}\label{SE}
S_E[\phi,\Phi]=S_0[\phi]+\ell S_\phi[\phi,\Phi]
+\ell S_{\phi\phi}[\phi,\Phi]+\ell S_\fff[\phi],
\end{equation}

\section{The quadratic terms}\label{sec:quad}

        We need to evaluate the path integral \eqref{infl} for the
influence functional, but the presence of a term \eqref{Lphiphiphi} in
$S_E$ which is cubic in $\phi$ prevents us from doing that in closed
form.  We might be led then to treat the problem perturbatively, with
the terms from \eqref{SE} first-order in $\ell$ providing a correction
to the answer obtained using the action $S_0$.  However, to zeroth
order in $\ell$, the theory if free and thus $e^{iW}=1$.  But if we
are to examine situations where $e^{iW}\ll1$, this is only possible if
the perturbative correction is also of order unity, which cannot
happen in a fully perturbative calculation.  We will see in
Sec.~\ref{ssec:breakdown} one set of circumstances where we can
perform some but not all of the calculations perturbatively and still
obtain $e^{iW}\ll1$.  In the meantime, we will consider those parts of
the action for which the integral \emph{can} be done
non-perturbatively, and add in the effects of the cubic term later,
using a perturbative treatment which pays careful attention to the
issues to be raised in Sec.~\ref{ssec:breakdown}.

        Since the parts of the action defined in \eqref{L0} and
\eqref{Lphiphi} are quadratic in $\phi$, it would be possible to do
the path integrals in \eqref{infl} explicitly if the action $S_E$
included only those terms.  Thus we turn our attention for the time
being to the modified influence functional
\begin{equation}
e^{iW_\off[\Phi,\Phi']}
\label{Woff}
=\int\D\phi\,\D\phi' 
\rho_\phi(\phi_i,\phi'_i)\delta(\phi'_f-\phi_f)
e^{i(S_\off[\phi,\Phi]-S_\off[\phi',\Phi'])}.
\end{equation}

        We will find an upper limit
\begin{equation}\tag{\ref{decTR}}
\abs{e^{iW_\off[\Phi,\Phi']}}
\le\left\{
        1+E^2[\Delta\Phi]
\right\}^{-1/4},
\end{equation}
on the absolute value of the influence functional in the absence of
the terms linear and cubic in the short-wavelength part $\phi$ of the
field.  We will show in Sec.~\ref{sec:full} and
Appendix~\ref{app:full} that restoring those terms does not
qualitatively change the limit \eqref{decTR}.

\subsection{A vector expression}\label{ssec:vec}

        The Lagrangian $L_\off$ can be written, using
$\phi_\M^*=\phi_{-\M}$, in the suggestive form
\begin{eqnarray}\label{Loff}
L_\off[\phi,\Phi]=L_0[\phi]+\ell L_{\phi\phi}[\phi,\Phi]
=\frac{1}{2}\int d^\Ddim\!k_1\, d^\Ddim\!k_2\,
\left\{\vphantom{\frac{1}{2}}\right.&&
\dot{\phi}_{1}^*[\delta^\Ddim\!(\M_1-\M_2)
-\ell\Phi_{1-2}]\dot{\phi}_{2}
-\frac{d}{dt}(\ell\phi_{1}^*\dot{\Phi}_{1-2}\phi_{2})
\\ \nonumber
&&-\phi_{1}^*
\bigl[
        \delta^\Ddim\!(\M_1-\M_2)\kay_{1}^2
        -\ell(k^2_{12}\Phi_{1-2}
        +\ddot{\Phi}_{1-2})
\bigr]
\phi_{2}
\left.\vphantom{\frac{1}{2}}
\right\}
\end{eqnarray}
where
\begin{equation}
k^2_{12}=-(\q_{1-2})\cdot(-\kay_{1})
-(-\kay_{1})\cdot\kay_{2}-\kay_{2}\cdot(\q_{1-2})
=\kay_{1}^2+\kay_{2}^2-\kay_{1}\cdot\kay_{2},\label{k2M}
\end{equation}
and we have defined
\begin{equation}
\q_{1\pm2}=\kay_1\pm\kay_2, \qquad \Phi_{1\pm2}=\Phi_{\q_{1\pm2}}.
\end{equation}
We would like to write \eqref{Loff} as a matrix expression in terms of
a vector which describes the short-wavelength modes $\{\phi_\M\}$.
However, the reality conditions
$\phi_{-\M}^{\mathrm{R}}=\phi_\M^{\mathrm{R}}$ and
$\phi_{-\M}^{\mathrm{I}}=-\phi_\M^{\mathrm{I}}$, which cause the
measure for the integral over independent modes to be
\begin{equation}\label{phimeas}
\Dend\phi\propto
\prod_{\M\in\mc{S}/2}\Dend\phi_\M^{\mathrm{R}}\,\Dend\phi_\M^{\mathrm{I}},
\end{equation}
necessitate some caution.  

        If we express the modes $\{\phi\}$ as a complex vector
$\phi_{\text{ex}}=\{\phi_\M|\M\in\mc{S}\}$, on a vector space we call,
with a slight abuse of notation, $\C^{\mc S}$, with inner product
\begin{equation}
w_{\text{ex}}^\dagger v_{\text{ex}}=\int\limits_{\mc{S}}
d^\Ddim\!k\, w_\M^* v_\M,
\end{equation}
the Lagrangian \eqref{Loff} can be written
\begin{equation}\label{Lex}
L_\off=
\frac{1}{2}\left[
\dot{\phi}_{\text{ex}}^\dagger m_{\text{ex}}\dot{\phi}_{\text{ex}}
-\phi_{\text{ex}}^\dagger \varpi_{\text{ex}}\phi_{\text{ex}}
+\frac{d}{dt}(\phi_{\text{ex}}^\dagger
\dot{m}_{\text{ex}}\phi_{\text{ex}})
\right]
\end{equation}
where $m_{\text{ex}}$ and $\varpi_{\text{ex}}$ are hermitian
matrices\footnote{or, more precisely, integral operators with these
kernels} acting on $\C^\mc{S}$ with the form
\begin{mathletters}\label{mex}
\begin{eqnarray}
(m_{\text{ex}})_{\M_1\M_2}&=&\delta^\Ddim\!(\M_1-\M_2)-\ell\Phi_{1-2}
\\
(\varpi_{\text{ex}})_{\M_1\M_2}&=&\delta^\Ddim\!(\M_1-\M_2)\kay_{1}^2
-\ell(k^2_{12}\Phi_{1-2}+\ddot{\Phi}_{1-2}).
\end{eqnarray}
\end{mathletters}
Unfortunately, the components of $\phi_{\text{ex}}$ represent twice as
many degrees of freedom as are integrated over in \eqref{phimeas}.
This means that a path integral over all the components would have
to include the factor
\begin{equation}
\prod_{\M\in\mc{S}/2}\delta\left[\phi_{-\M}-\phi_\M^*\right].
\end{equation}

\begin{figure}
\begin{center}
\epsfig{file=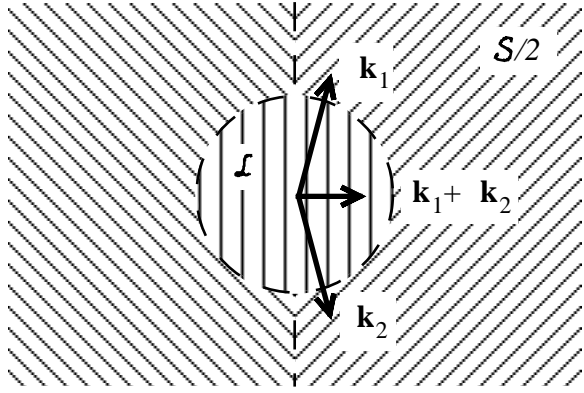}
\end{center}
\caption[The addition of momenta $\M_1,\M_2\in\mc{S}/2$ can produce
$\M_1+\M_2\in\mc{L}$] {The addition of momenta $\M_1,\M_2\in\mc{S}/2$
can produce $\M_1+\M_2\in\mc{L}$.  The long-wavelength (low-momentum)
region $\mc{L}$ is shaded vertically.  The short-wavelength
(high-momentum) region $\mc{S}$ is shaded diagonally in one direction
or the other.  The right half of $\mc{S}$, shaded diagonally up and to
the right, is $\mc{S}/2$.  In $D>1$, we see that it is possible to add
two ``large'' momenta on the right side of the origin
($\M_1,\M_2\in\mc{S}/2$) to get a ``small'' momentum
($\M_1+\M_2\in\mc{L}$).}
\label{fig:LS2}
\end{figure}

        On the other hand, a complex vector
$\phi_{+}=\{\phi_\M|\M\in\mc{S}/2\}$ in a space $\C^{\mc{S}/2}$ with
inner product
\begin{equation}
w_{+}^\dagger v_{+}=\int\limits_{\mc{S}/2}
d^\Ddim\!k\, w_\M^* v_\M,
\end{equation}
would completely specify the unique modes of
$\phi$, but \eqref{Loff} is not conveniently expressed in terms of
$\phi_{+}$.  To see this, consider the velocity term
\begin{eqnarray}
\frac{1}{2}\int\limits_{\mc{S}} d^\Ddim\!k_1\,&& d^\Ddim\!k_2\,
\dot{\phi}_{1}^*[\delta^\Ddim\!(\M_1-\M_2)-\ell\Phi_{1-2}]
\dot{\phi}_{2}
\\ \nonumber
&&=\frac{1}{2}\int\limits_{\mc{S}/2} d^\Ddim\!k_1\, d^\Ddim\!k_2\,
\Bigl\{
\dot{\phi}_{1}^*\left[\delta^\Ddim\!(\M_1-\M_2)-\ell\Phi_{1-2}\right]
\dot{\phi}_{2}
+\dot{\phi}_{1}\left[\delta^\Ddim\!(\M_1-\M_2)-\ell\Phi_{2-1}\right]
\dot{\phi}_{2}^*
-\dot{\phi}_{1}^*\ell\Phi_{1+2}\dot{\phi}_{2}^*
-\dot{\phi}_{1}\ell\Phi_{1+2}^*\dot{\phi}_{2}
\Bigr\}
\end{eqnarray}
Although the first two terms can be written as
\begin{equation}
\int\limits_{\mc{S}/2} d^\Ddim\!k_1\, d^\Ddim\!k_2\,
\dot{\phi}_{1}^*\left[\delta^\Ddim\!(\M_1-\M_2)-\ell\Phi_{1-2}\right]
\dot{\phi}_{2}
=\phi_{+}^\dagger m_{+}\phi_{+}
\end{equation}
(where $m_{+}$ is the restriction of $m_{\text{ex}}$ to
$\C^{\mc{S}/2}$), the last two give
\begin{equation}
\Real\left(
        \int\limits_{\mc{S}/2} d^\Ddim\!k_1\, d^\Ddim\!k_2\,
        \dot{\phi}_{1}
        \left[\delta^\Ddim\!(\M_1-\M_2)-\ell\Phi_{1+2}^*\right]
        \dot{\phi}_{2}
\right),
\end{equation}
which cannot be written in terms of the complex vector $\phi_{+}$ and
its adjoint $\phi_{+}^\dagger$ without using the transpose
$\trans{\phi}_{+}$ or the complex conjugate $\phi_{+}^*$.  If
$D=1$, this is not a problem, since $\M_1,\M_2\in\mc{S}/2$
implies $\M_1+\M_2\notin\mc{L}$ and hence $\Phi_{1+2}=0$.
However, it is possible in $D>1$ to have $\M_1+\M_2\in\mc{L}$
even when $\M_1,\M_2\in\mc{S}/2$, as illustrated in Fig.~\ref{fig:LS2}.

        The most useful approach is to define a real vector
\begin{equation}\label{phivec}
\phi=\{\sqrt{2}\,\phi_\M^{\mathrm{R}},\sqrt{2}\,\phi_\M^{\mathrm{I}}
|\M\in\mc{S}/2\}
\end{equation}
in the space $\R^{\mc{S}/2}\otimes\R^{\mc{S}/2}
=\R^{\mc{S}/2\oplus\mc{S}/2}$ with inner product
\begin{equation}
\trans{w}v=2\int\limits_{\mc{S}/2}d^\Ddim\!k\, 
(w_\M^{\mathrm{R}}v_\M^{\mathrm{R}}
+w_\M^{\mathrm{I}}v_\M^{\mathrm{I}}).
\end{equation}
A straightforward calculation shows that it is possible to write
\eqref{Lex} as
\begin{equation}\label{Lexvec}
L_\off=
\frac{1}{2}\left[
\trans{\dot{\phi}} m\dot{\phi}
-\trans{\phi}\varpi\phi
+\frac{d}{dt}(\trans{\phi}\dot{m}\phi)
\right],
\end{equation}
where $m$ and $\varpi$ are real symmetric matrices on
$\R^{\mc{S}/2\oplus\mc{S}/2}$, given by
\begin{eqnarray}
\label{mdef}
m&=&
\left(\begin{array}{cc}
\{\delta^\Ddim\!(\M_1-\M_2)
-\ell(\Phi_{1-2}^{\mathrm{R}}+\Phi_{1+2}^{\mathrm{R}})\} &
\{-\ell(-\Phi_{1-2}^{\mathrm{I}}+\Phi_{1+2}^{\mathrm{I}})\} \\
\{-\ell(\Phi_{1-2}^{\mathrm{I}}+\Phi_{1+2}^{\mathrm{I}})\} &
\{\delta^\Ddim\!(\M_1-\M_2)
-\ell(\Phi_{1-2}^{\mathrm{R}}-\Phi_{1+2}^{\mathrm{R}})\}
\end{array}\right)
\\
\label{varpidef}
\varpi&=&
\left(\begin{array}{cc}
\{\varpi^{\mathsf{UL}}_{\M_1\M_2}\} & \{\varpi^{\mathsf{UR}}_{\M_1\M_2}\}
\\
\{\varpi^{\mathsf{LL}}_{\M_1\M_2}\} & \{\varpi^{\mathsf{LR}}_{\M_1\M_2}\}
\end{array}\right)
\end{eqnarray}
and
\addtocounter{equation}{-1}
\begin{mathletters}
\begin{eqnarray}
\varpi^{\mathsf{UL}}_{\M_1\M_2}&=&
\delta^\Ddim\!(\M_1-\M_2)k_{1}^2
-\ell(\Phi_{1-2}^{\mathrm{R}}k^2_{12}
        +\Phi_{1+2}^{\mathrm{R}}k^2_{1,-2}
        +\ddot{\Phi}_{1-2}^{\mathrm{R}}
        +\ddot{\Phi}_{1+2}^{\mathrm{R}}
)\\
\varpi^{\mathsf{UR}}_{\M_1\M_2}&=&
-\ell(-\Phi_{1-2}^{\mathrm{I}}k^2_{12}
        +\Phi_{1+2}^{\mathrm{I}}k^2_{1,-2}
        -\ddot{\Phi}_{1-2}^{\mathrm{I}}
        +\ddot{\Phi}_{1+2}^{\mathrm{I}}
)\\
\varpi^{\mathsf{LL}}_{\M_1\M_2}&=&
-\ell(\Phi_{1-2}^{\mathrm{I}}k^2_{12}
        +\Phi_{1+2}^{\mathrm{I}}k^2_{1,-2}
        +\ddot{\Phi}_{1-2}^{\mathrm{I}}
        +\ddot{\Phi}_{1+2}^{\mathrm{I}}
)\\
\varpi^{\mathsf{LR}}_{\M_1\M_2}&=&
\delta^\Ddim\!(\M_1-\M_2)k_{1}^2
-\ell(\Phi_{1-2}^{\mathrm{R}}k^2_{12}
        -\Phi_{1+2}^{\mathrm{R}}k^2_{1,-2}
        +\ddot{\Phi}_{1-2}^{\mathrm{R}}
        -\ddot{\Phi}_{1+2}^{\mathrm{R}}
);
\end{eqnarray}
\end{mathletters}
where $\phi$ is written as
\begin{equation}\label{phidef}
\phi=\sqrt{2}
\left(\begin{array}{c}
\phi^{\mathrm{R}}_{+}\\ \phi^{\mathrm{I}}_{+}
\end{array}\right)
=\sqrt{2}
\left(\begin{array}{c}
\{\phi_\M^{\mathrm{R}}\}\\ \{\phi_\M^{\mathrm{I}}\}
\end{array}\right).
\end{equation}

\subsection{The propagator}

        We now have a workable vector expression for the path integral
\eqref{Woff}:
\begin{eqnarray}
e^{iW_\off[\Phi,\Phi']}
&=&\int\D\phi\,\D\phi' 
\rho_\phi(\phi_i,\phi'_i)\delta(\phi'_f-\phi_f)
e^{i(S_\off[\phi,\Phi]-S_\off[\phi',\Phi'])}
\nonumber\\
&=&\int\Dend\phi_i\,\Dend\phi'_i\,\Dend\phi_f\,
\rho_\phi(\phi_i,\phi'_i)
\mc{K}_\off(\phi_f|\phi_i;\Phi]
\mc{K}^*_\off(\phi_f|\phi'_i;\Phi']\label{WoffK}
\end{eqnarray}
where
\begin{equation}
\mc{K}_\off(\phi_f|\phi_i;\Phi]
=\int\limits_{\phi_f\phi_i}\D\phi\,e^{iS_\off[\phi,\Phi]}
\end{equation}
is the propagator for the quadratic action.  It is useful to write
\begin{equation}\label{Koff}
\mc{K}_\off(\phi_f|\phi_i;\Phi]
=e^{\frac{i}{2}(\trans{\phi_f}\dot{m}_f\phi_f-\trans{\phi_i}\dot{m}_i\phi_i)}
\,K\!\left(\phi_f\,\frac{T}{2}\right.\left|\phi_i\,-\frac{T}{2}\right)
\end{equation}
where
\begin{equation}
K(\phi_f t_f|\phi_i t_i)
=\int\limits_{\phi_f\phi_i}\D\phi
\,
e^{
        \frac{i}{2}\int_{t_i}^{t_f} dt\,[\trans{\dot{\phi}(t)}m(t)\phi(t)
        -\trans{\phi(t)}\varpi(t)\phi(t)]
}
\end{equation}
is the propagator  for a simple harmonic
oscillator with time-dependent matrices $m(t)$ and $\varpi(t)$ in
place of $m$ and $m\omega^2$.  [The dependence on $\Phi$ is now
implicit in the time dependence of $m(t)$ and $\varpi(t)$, given by
(\ref{mdef}--\ref{varpidef}).]

This propagator can be found explicitly to be\cite{dodonov}
\begin{equation}\label{prp}
K(\phi_f t_f|\phi_i t_i)
=\frac{1}{\sqrt{\det\lbp 2\pi i\Crune(t_f|t_i)\rbp}}
\exp\left[
        \frac{i}{2}
\trans{
  \left(\begin{array}{c}
        \phi_f\\ \phi_i
  \end{array}\right)
}
\left(\begin{array}{cc}
        \Crune^{-1}(t_f|t_i)\Brune(t_f|t_i)     & -\Crune^{-1}(t_f|t_i)  \\
        -\trans{\Crune^{-1}(t_f|t_i)}   & \Arune(t_f|t_i)\Crune^{-1}(t_f|t_i)
\end{array}\right)
\left(\begin{array}{c}
        \phi_f\\ \phi_i
\end{array}\right)
\right],
\end{equation}
where
\begin{mathletters}\label{solns}
\begin{eqnarray}
\label{Bsoln}
\Brune(t_f|t_i)&=&\sum_{n=0}^\infty\left(
        \prod_{k=1}^n
        \int_{t_i}^{\tilde{t}_{k-1}}dt_k
        \int_{t_i}^{t_k}d\tilde{t}_k
\right)
\prod_{k=n}^1\left[
        -m^{-1}(\tilde{t}_k)\varpi(t_k)
\right]
\\
\label{Csoln}
\Crune(t_f|t_i)&=&-\frac{d\Brune(t_f|t_i)}{dt_f}\varpi^{-1}(t_f)
\\
\label{Asoln}
\Arune(t_f|t_i)&=&-m(t_i)\frac{d\Crune(t_f|t_i)}{dt_i}
\end{eqnarray}
\end{mathletters}
are the solutions to
\begin{mathletters}\label{dtb}
\begin{eqnarray}
\frac{d\Arune(t_f|t_i)}{dt_f}
&=&\bigl[\Arune(t_f|t_i)\Crune^{-1}(t_f|t_i)\Brune(t_f|t_i)
-\trans{\Crune^{-1}(t_f|t_i)}\bigr]m^{-1}(t_f)
\\
\frac{d\Brune(t_f|t_i)}{dt_f}&=&-\Crune(t_f|t_i)\varpi(t_f)
\\
\frac{d\Crune(t_f|t_i)}{dt_f}&=&\Brune(t_f|t_i)m^{-1}(t_f),
\end{eqnarray}
\end{mathletters}
or 
\begin{mathletters}\label{dta}
\begin{eqnarray}
\frac{d\Arune(t_f|t_i)}{dt_i}&=&\varpi(t_i)\Crune(t_f|t_i)
\\
\frac{d\Brune(t_f|t_i)}{dt_i}
&=&-m^{-1}(t_i)\bigl[\Arune(t_f|t_i)\Crune^{-1}(t_f|t_i)\Brune(t_f|t_i)
-\trans{\Crune^{-1}(t_f|t_i)}\bigr]
\\
\frac{d\Crune(t_f|t_i)}{dt_i}&=&-m^{-1}(t_i)\Arune(t_f|t_i),
\end{eqnarray}
\end{mathletters}
with the initial conditions $\Arune(t|t)=1=\Brune(t|t)$ and
$\Crune(t|t)=0$.

These exact expressions are expanded to first order in $\ell$ in
Sec.~\ref{ssec:pertrune} of Appendix~\ref{app:pert}, using the values
of $m(t)$ and $\varpi(t)$ given by \eqref{mdef} and \eqref{varpidef},
respectively.

        Given the expression \eqref{prp} for the time-dependent
propagator, \eqref{Koff} becomes
\begin{equation}\label{Koffrm}
\mc{K}_\off(\phi_f|\phi_i;\Phi]
=\frac{1}{\sqrt{\det(2\pi i\Crune[\Phi])}}
\exp\left[
        \frac{i}{2}
\trans{
  \left(\begin{array}{c}
        \phi_f\\ \phi_i
  \end{array}\right)
}
\left(\begin{array}{cccc}
        B[\Phi] & -C[\Phi] \\
        -\trans{C[\Phi]} & A[\Phi]
\end{array}\right)
\left(\begin{array}{c}
        \phi_f\\ \phi_i
\end{array}\right)
\right],
\end{equation}
where
\begin{mathletters}\label{ABC}
\begin{eqnarray}
A[\Phi]&=&\Arune\left(\frac{T}{2}\right.\left|-\frac{T}{2}\right)
\Crune^{-1}\left(\frac{T}{2}\right.\left|-\frac{T}{2}\right)
-\dot{m}\left(-\frac{T}{2}\right)\label{APdef}
\\
B[\Phi]&=&\Crune^{-1}\left(\frac{T}{2}\right.\left|-\frac{T}{2}\right)
\Brune\left(\frac{T}{2}\right.\left|-\frac{T}{2}\right)
+\dot{m}\left(\frac{T}{2}\right)\label{BPdef}
\\
C[\Phi]&=&\Crune^{-1}\left(\frac{T}{2}\right.\left|-\frac{T}{2}\right).
\label{CPdef}
\end{eqnarray}
\end{mathletters}
This means that \eqref{WoffK} becomes
\begin{equation}\label{WoffA}
e^{iW_\off[\Phi,\Phi']}
=\int \frac{\Dend\phi_i\,\Dend\phi'_i\,\Dend\phi_f\,\rho_\phi(\phi_i,\phi'_i)}
{\sqrt{\det(2\pi\Crune[\Phi])\det(2\pi\Crune[\Phi'])}}
\exp\left[
        \frac{i}{2}
\trans{
  \left(\begin{array}{c}
        \phi_f \\ \phi_i \\ \phi'_i
  \end{array}\right)
}
\left(\begin{array}{cccc}
        B[\Phi]-B[\Phi'] & -C[\Phi] & C[\Phi'] \\
        -\trans{C[\Phi]} & A[\Phi] & 0 \\
        \trans{C[\Phi']} & 0 & -A[\Phi']
\end{array}\right)
\left(\begin{array}{c}
        \phi_f \\ \phi_i \\ \phi'_i
\end{array}\right)
\right]
\end{equation}

\subsection{The initial state}

        Before we can perform the integrals over the endpoints
$\phi_i$, $\phi'_i$ and $\phi_f$ of the SWM paths which remain in the
expression \eqref{WoffA} for the influence functional, we need to
specify the the initial state $\rho_\phi$ of the SWM environment.  One
simple choice is a thermal state with temperature
$1/k_{{\textsc{b}}}\beta$.  This is physically reasonable if we
consider the main source of such an environment to be, for instance,
the primordial graviton background.  \cite{K&T}

        The density matrix for this state is given as an operator by
$\widehat{\rho}\propto e^{-\beta\widehat{H}}$.  Using the full
Hamiltonian corresponding to the action \eqref{SE} would couple the
short- and long-wavelength modes, preventing the separation
\eqref{rhosep} of the initial state.  So instead we use the zero-order
non-interacting action $S_0$, which gives the thermal density matrix
for a simple harmonic oscillator of frequency $\Oo=\left(
\begin{array}{cc}\diag\{k\}&0\\0&\diag\{k\}\end{array}
\right)$ and unit mass:
\begin{equation}
\rho_\phi(\phi_i,\phi'_i)
\propto\exp\left[
-\frac{1}{2}
\trans{
  \left(\begin{array}{c}
        \phi_i \\ \phi'_i
  \end{array}\right)
}
\left(\begin{array}{cc}
        \frac{\Oo}{\tanh\Oo\beta}
                &       -\frac{\Oo}{\sinh\Oo\beta}\\
        -\frac{\Oo}{\sinh\Oo\beta}
                &       \frac{\Oo}{\tanh\Oo\beta}
\end{array}\right)
\left(\begin{array}{c}
        \phi_i \\ \phi'_i
\end{array}\right)
\right]
\label{rhodef}
\end{equation}

Equation~\eqref{Koffrm} is simplified if we express it in terms of
$\overline{\phi_i}=\frac{\phi_i+\phi'_i}{2}$ and
$\Delta\phi_i=\phi_i-\phi'_i$ using
\begin{equation}\label{bardef}
\left(\begin{array}{c}
        \phi_i \\ \phi'_i
\end{array}\right)
=\left(\begin{array}{cccc} 
1 & 1 \\ 1 & -1
\end{array}\right)
\left(\begin{array}{c}
        \overline{\phi_i} \\ \Delta\phi_i/2
\end{array}\right).
\end{equation}
Both $\frac{\Oo}{\tanh\Oo\beta}
-\frac{\Oo}{\sinh\Oo\beta}$ and
$\frac{\Oo}{\tanh\Oo\beta}
+\frac{\Oo}{\sinh\Oo\beta}$ can be expressed in terms of
\begin{equation}
\mc{V}(\Oo)
=\frac{2}{\Oo}\frac{\cosh\Oo\beta-1}{\sinh\Oo\beta}
=\frac{2}{\Oo}\frac{\sinh\Oo\beta}{\cosh\Oo\beta+1}
=\frac{2}{\Oo}\sqrt{\frac{\cosh\Oo\beta-1}{\cosh\Oo\beta+1}}
=\frac{2}{\Oo}\tanh\frac{\Oo\beta}{2}
\end{equation}
to give
\begin{equation}
\rho_\phi(\phi_i,\phi'_i)
\propto\exp\left[
-\frac{1}{2}
\trans{
  \left(\begin{array}{c}
        \overline{\phi_i} \\ \Delta\phi_i/2
  \end{array}\right)
}
\left(\begin{array}{cccc}
        \Oo^2\mc{V}(\Oo) & 0 \\ 0 & \mc{V}^{-1}(\Oo)
\end{array}\right)
\left(\begin{array}{c}
        \overline{\phi_i} \\ \Delta\phi_i/2
\end{array}\right)
\right];
\end{equation}
if we also define
\begin{mathletters}\label{Apm}
\begin{eqnarray}
\mc{A}_{\pm}&=&A[\Phi]\pm A[\Phi']
\\
\mc{B}_{\pm}&=&B[\Phi]\pm B[\Phi']
\\
\mc{C}_{\pm}&=&C[\Phi]\pm C[\Phi'],
\end{eqnarray}
\end{mathletters}
\eqref{WoffA} becomes
\begin{eqnarray}
\label{Woffdet}
e^{iW_\off[\Phi,\Phi']}
&\propto&\int\frac{\Dend\overline{\phi_i}\,\Dend\Delta\phi_i\,\Dend\phi_f}
{\sqrt{\det(2\pi\Crune[\Phi])\det(2\pi\Crune[\Phi'])}}
\exp\left[
        -\frac{1}{2}
\trans{
  \left(\begin{array}{c}
        \phi_f \\ \overline{\phi_i} \\ \Delta\phi_i/2
  \end{array}\right)
}
\mc{M}
\left(\begin{array}{c}
        \phi_f \\ \overline{\phi_i} \\ \Delta\phi_i/2
\end{array}\right)
\right]
\\\nonumber
&=&
\left\{\det(2\pi\Crune[\Phi])\det(2\pi\Crune[\Phi'])
\det\left(\frac{\mc{M}}{2\pi}\right)
\right\}^{-1/2},
\end{eqnarray}
so that the calculation of $e^{iW_\off}$ is reduced to the evaluation
of the determinants of $\Crune$ and
\begin{equation}\label{Mdef}
\mc{M}
=\left(\begin{array}{cccc}
-i\mc{B}_{-} & i\mc{C}_{-} & i\mc{C}_{+} \\
i\trans{\mc{C}_{-}} & \Oo^2\mc{V}(\Oo)-i\mc{A}_{-} & -i\mc{A}_{+} \\
i\trans{\mc{C}_{+}} & -i\mc{A}_{+} & 4\mc{V}^{-1}(\Oo)-i\mc{A}_{-}
\end{array}\right).
\end{equation}

\subsection{Controlling the breakdown of perturbation theory}
\label{ssec:breakdown}

        Up to this point, the treatment has been completely
non-perturbative\footnote{Although the action \eqref{toyact} was of
course obtained using perturbative considerations.} and we have
successfully performed all of the integrations over $\phi$ contained
in \eqref{Woff}.  However, the expression obtained depends on the
functionals $A[\Phi]$, $B[\Phi]$ and $C[\Phi]$, which, while they can
be written exactly in terms of \eqref{ABC} and \eqref{solns}, are best
understood using the expansions in powers of $\ell$ from
Appendix~\ref{app:pert}.  If we are going to begin to expand in powers
of $\ell$, however, we need to address the issue of how to obtain
decoherence via a partially perturbative calculation.

        As alluded to before, if we try to expand the influence
functional $e^{iW}$ defined by \eqref{infl} in powers of $\ell$, we
note that as the zero-order term in $S_E[\phi,\Phi]$ is just
$S_0[\phi]$ [{i.e.,} the ``system'' and ``environment'' are
decoupled to zeroth order; {cf.}\ \eqref{SE}],
\begin{equation}\label{W0}
\left(e^{iW[\Phi,\Phi']}\right)_0
=\int\D\phi\,\D\phi' 
\rho_\phi(\phi_i,\phi'_i)\delta(\phi'_f-\phi_f)
e^{i(S_0[\phi]-S_0[\phi'])}
=\Tr\left[e^{-iT\widehat{H}}\rho_\phi e^{iT\widehat{H}}\right]=1.
\end{equation}
Perturbatively, then, we would conclude $e^{iW}=1+\mc{O}(\ell)$.  The
problem is that for the influence phase to be effective at producing
decoherence, we need $e^{iW[\Phi,\Phi']}\ll 1$ for $\Phi$ and $\Phi'$
sufficiently different.  This can only be possible if the perturbative
analysis breaks down somehow.  In this sense, \emph{decoherence is an
inherently nonperturbative phenomenon.}

        We will focus on one scenario in which perturbation theory
cannot be applied universally, but it is easy to keep track of which
seemingly negligible terms must be retained.  The basic idea can be
expressed simply.  Consider a quantity
\begin{equation}
F(a,b)=I(a)+\frac{A(a)}{b}
\end{equation}
which depends on two parameters $a$ and $b$, and suppose the functions
$I(a)$ and $A(a)$ have expansions
\begin{eqnarray}
I(a)&=&1+\mc{O}(a)\\
A(a)&=&a+\mc{O}(a^2).
\end{eqnarray}
If we consider only the behavior of $F$ as perturbative expansion in
$a$, we would conclude that
\begin{equation}
F(a,b)=1+\mc{O}(a).
\end{equation}
If we were looking for cases where $F\gg1$, we would conclude that
they do not exist in the perturbative regime.  However, if $b$ is also
small, have to be more careful about
\begin{equation}
F(a,b)=1+\frac{a}{b}+\mc{O}\left(a,\frac{a^2}{b}\right).
\end{equation}
Unless $a\ll b$, we cannot neglect $a/b$ relative to 1.  However, we
can still neglect the $\mc{O}(a)$ terms [from expansion of $I(a)$]
relative to 1, and the $\mc{O}(a^2/b)$ terms [from further expansion
of $A(a)$] relative to $a/b$.  Thus even when it is valid to use
perturbative (in fact, lowest order) expressions for $I(a)$ and
$A(a)$, it is still possible to have $F\gg1$ (when $b\ll a$), since it
is not valid to expand $F(a,b)$ simply in powers of $a$.

        In the current problem, where the role of $F$ is played by
$1/\abs{e^{iW}}$, and the small parameter corresponding to $a$ is
$\ell$, the role of the additional parameter $b$ can be played by
$\beta$.  If the temperature $\beta^{-1}$ is high enough, there will
be some modes in $\mc{S}$ for which $\mc{V}(k)=\frac{2}{k}\frac{\sinh
k\beta}{\cosh k\beta+1}\rightarrow\beta$, and the $\mc{O}(\beta)$
terms like $\mc{C}_{+}\alpha^{-1}\trans{\mc{C}_{+}}$ may become
smaller than $\mc{O}(\ell)$ terms like $\mc{B}_{-}$. At that point, if
the $\mc{O}(\ell)$ correction to $e^{iW}$ is also
$\mc{O}(\beta^{-1})$, it can cause perturbation theory to break down.
We keep a handle on this breakdown by neglecting $\mc{O}(\ell)$ terms
only when they are not compared to potentially $\mc{O}(\beta)$ terms.

\subsection{Evaluation of the influence functional}\label{SSEC:EVAL}

 Since the matrix $A[\Phi]$ can be expanded (see Sec.~\ref{ssec:pert}
of Appendix~\ref{app:pert}) as $A[\Phi]=A_0+\ell
A_1[\Phi]+\mc{O}(\ell^2)$, where $A_1[\Phi]$ is a linear functional of
its argument, we have $\mc{A_{+}}=2A_0+\mc{O}(\ell)$ and
$\mc{A_{-}}=\ell A_1[\Delta\Phi]+\mc{O}(\ell^2)$ (where
$\Delta\Phi=\Phi-\Phi'$), with similar expressions holding for
$\mc{B}_{\pm}$ and $\mc{C}_{\pm}$.  This means the sub-matrices of
$\mc{M}$ are of the following order:
\begin{equation}
\mc{M}
=\left(\begin{array}{cccc}
\mc{O}(\ell) & \mc{O}(\ell) & \mc{O}(1) \\
\mc{O}(\ell) & \Oo^2\mc{V}(\Oo)+\mc{O}(\ell) & \mc{O}(1) \\
\mc{O}(1) & \mc{O}(1) & 4\mc{V}^{-1}(\Oo)+\mc{O}(\ell)
\end{array}\right).
\end{equation}
Given the relation
\begin{equation}
4\mc{V}^{-1}(\Oo)=
2\Oo\sqrt{\frac{\cosh\Oo\beta+1}{\cosh\Oo\beta-1}}\ge 2\Oo 
\ge2\Oo\sqrt{\frac{\cosh\Oo\beta-1}{\cosh\Oo\beta+1}}=\Oo^2\mc{V}(\Oo)
\end{equation}
we see that $\alpha=4\mc{V}^{-1}(\Oo)-i\mc{A}_{-}$ is the largest
of the sub-matrices on the diagonal, and is no smaller than $\mc{O}(1)$.
Thus we partially diagonalize $\mc{M}$ about it to get
\begin{eqnarray}
\wt{\mc{M}}
&=&
\left(\begin{array}{ccc}
1 & 0 & -i\mc{C}_{+}\alpha^{-1} \\
0 & 1 & i\mc{A}_{+}\alpha^{-1} \\
0 & 0 & 1
\end{array}\right)
\mc{M}\left(\begin{array}{ccc}
1 & 0 & 0 \\
0 & 1 & 0 \\
-i\alpha^{-1}\trans{\mc{C}_{+}} & i\alpha^{-1}\mc{A}_{+} & 1
\end{array}\right)
\nonumber\\
&=&\left(\begin{array}{ccc}
\mc{C}_{+}\alpha^{-1}\trans{\mc{C}_{+}}-i\mc{B}_{-}  &
   -\mc{C}_{+}\alpha^{-1}\mc{A}_{+}+i\mc{C}_{-} &       0 \\
-\mc{A}_{+}\alpha^{-1}\trans{\mc{C}_{+}}+i\trans{\mc{C}_{-}}  &
   \mc{A}_{+}\alpha^{-1}\mc{A}_{+}+\Oo^2\mc{V}(\Oo)-i\mc{A}_{-} & 0 \\
0       &       0       &       \alpha
\end{array}\right)\label{Mtilde}.
\end{eqnarray}

        Using the approximation of Sec.~\ref{ssec:breakdown}, we have
\begin{eqnarray}
\wt{\mc{M}}
&=&\left(\begin{array}{cccc}
\mc{C}_{+}\frac{\mc{V}(\Oo)}{4}\trans{\mc{C}_{+}}-i\mc{B}_{-}  &
  -\mc{C}_{+}\frac{\mc{V}(\Oo)}{4}\mc{A}_{+}+i\mc{C}_{-} \\
-\mc{A}_{+}\frac{\mc{V}(\Oo)}{4}\trans{\mc{C}_{+}}+i\trans{\mc{C}_{-}}  &
  \mc{A}_{+}\frac{\mc{V}(\Oo)}{4}\mc{A}_{+}+\Oo^2\mc{V}(\Oo)-i\mc{A}_{-}
\end{array}\right)
\oplus
4\mc{V}^{-1}(\Oo)
\nonumber\\
&=&\left(\begin{array}{cccc}
\frac{\Oo^2\mc{V}(\Oo)}{\sin^2\Oo T}-i\ell B_1[\Delta\Phi]
& -\frac{\Oo^2\mc{V}(\Oo)\cos\Oo T}{\sin^2\Oo T}+i\ell C_1[\Delta\Phi] \\
-\frac{\Oo^2\mc{V}(\Oo)\cos\Oo T}{\sin^2\Oo T}+i\trans{\ell C_1[\Delta\Phi]}
& \frac{\Oo^2\mc{V}(\Oo)}{\sin^2\Oo T}-i\ell A_1[\Delta\Phi]
\end{array}\right)
\oplus
4\mc{V}^{-1}(\Oo).
\end{eqnarray}
Noting that the matrices used to perform the diagonalization in
\eqref{Mtilde} have unit determinant, we have
\begin{equation}
\det\mc{M}=\det\wt{\mc{M}}=\det\lbp4\mc{V}^{-1}(\Oo)
\rbp\det(\aleph_0-i\ell\aleph_1[\Delta\Phi])
\propto\det
\left(
        1-i\ell\aleph_0^{-1/2}\aleph_1[\Delta\Phi]\aleph_0^{-1/2}
\right)
\end{equation}
where
\begin{mathletters}\label{aleph}
\begin{eqnarray}
&&\aleph_0=\frac{\Oo^2\mc{V}(\Oo)}{\sin^2\Oo T}
\left(\begin{array}{cc}
1 & -\cos\Oo T \\ -\cos\Oo T & 1
\end{array}\right)
\label{alepho}
\\
\label{alephi}
&&\aleph_1[\Delta\Phi]=
\left(\begin{array}{cc}
B_1[\Delta\Phi] & -C_1[\Delta\Phi] \\
-\trans{C_1[\Delta\Phi]} & A_1[\Delta\Phi]
\end{array}\right).
\end{eqnarray}
\end{mathletters}

        Now, $e^{i\Real W}$ is simply a phase multiplying the
decoherence functional \eqref{decinfl}; the part which can actually
make the off-diagonal components of $D[\Phi,\Phi']$ small is
$e^{-\Imag W}=\abs{e^{iW}}$.  Noting from \eqref{rune0C}
that the factors of $\det\Crune$
in \eqref{Woffdet} give, to lowest order in $\ell$, the
$\Phi$-independent values $\det\left(\frac{\sin\Oo T}{\Oo}\right)$, we
have [{cf.} \eqref{Woffdet}]
\begin{equation}
\abs{e^{iW_\off[\Phi,\Phi']}}\propto
\left[\det\left(\mc{M}^\dagger\mc{M}\right)\right]^{-1/4}
\propto\left\{
        \det\left(
                1+\ell^2\aleph_0^{-1/2}\aleph_1[\Delta\Phi]
                \aleph_0^{-1}\aleph_1[\Delta\Phi]\aleph_0^{-1/2}
        \right)
\right\}^{-1/4}.
\end{equation}
The normalization is set by \eqref{W0}, and in fact
\begin{equation}
\abs{e^{iW_\off[\Phi,\Phi']}}
=\left\{
        \det\left(
                1+\ell^2\aleph_0^{-1/2}\aleph_1[\Delta\Phi]
                \aleph_0^{-1}\aleph_1[\Delta\Phi]\aleph_0^{-1/2}
        \right)
\right\}^{-1/4}.
\end{equation}
For any positive matrix $a^2$, a straightforward analysis in the
diagonal basis shows $\det(1+a^2)\ge 1+\Tr a^2$, so
\begin{equation}\label{decTR}
\abs{e^{iW_\off[\Phi,\Phi']}}
\le\left\{
        1+E^2[\Delta\Phi]
\right\}^{-1/4},
\end{equation}
where
\begin{equation}
E^2[\Delta\Phi]=
\Tr\left(\ell\aleph_0^{-1}\aleph_1[\Delta\Phi]\right)^2
\end{equation}

        The magnitude of the influence functional \eqref{decTR} will
be small when
$E^2[\Delta\Phi]$ is large.
This is calculated in Appendix~\ref{app:aleph} and found to be
\begin{eqnarray}
E^2[\Delta\Phi]=\int\limits_{k_1,k_2>k_c}
\frac{d^\Ddim\!k_1\,d^\Ddim\!k_2\,\Theta(k_c-q)}
{4\mc{V}(k_1)\mc{V}(k_2)}&&
\left\{
\vphantom{\abs{\int_{-T/2}^{T/2}}^2}
\right.
        \abs{\sin^2\frac{\theta_{12}}{2}
        \int_{-T/2}^{T/2}dt\,\ell\Delta\Phi_\q(t)e^{ik_{-}t}
        -i\frac{k_{-}}{k_1k_2}
        \left[e^{i2k_{-}t}\ell\Delta\Phi_\q(t)
        \right]_{-T/2}^{T/2}}^2
\nonumber\\
        &&+\abs{\sin^2\frac{\theta_{12}}{2}
        \int_{-T/2}^{T/2}dt\,\ell\Delta\Phi_\q(t)e^{-ik_{-}t}
        +i\frac{k_{-}}{k_1k_2}
        \left[e^{-i2k_{-}t}\ell\Delta\Phi_\q(t)
        \right]_{-T/2}^{T/2}}^2
\nonumber\\
        &&+\abs{\cos^2\frac{\theta_{12}}{2}
        \int_{-T/2}^{T/2}dt\,\ell\Delta\Phi_\q(t)e^{ik_{+}t}
        +i\frac{k_{+}}{k_1k_2}
        \left[e^{i2k_{+}t}\ell\Delta\Phi_\q(t)
        \right]_{-T/2}^{T/2}}^2
\label{TRcont}\\ \nonumber
        &&+\abs{\cos^2\frac{\theta_{12}}{2}
        \int_{-T/2}^{T/2}dt\,\ell\Delta\Phi_\q(t)e^{-ik_{+}t}
        -i\frac{k_{+}}{k_1k_2}
        \left[e^{-i2k_{+}t}\ell\Delta\Phi_\q(t)
        \right]_{-T/2}^{T/2}}^2
\left.
\vphantom{\abs{\int_{-T/2}^{T/2}}^2}
\right\},
\end{eqnarray}
where $\q=\kay_1-\kay_2$, $k_{\pm}=k_1\pm k_2$, and
$\cos\theta_{12}=\frac{\kay_1\cdot\kay_2}{k_1 k_2}$.

\subsubsection{Specializing to D=3}\label{sssec:3D}

        If there are three spatial dimensions, the integration in
\eqref{TRcont} is over the six components of $\kay_1$ and $\kay_2$.
The integrand, however, is expressed in terms of the three components
of $\q$ and the two amplitudes $k_1$ and $k_2$ (or equivalently,
$k_{\pm}$).  There is also a dependence on
$\cos\theta_{12}=\frac{\kay_1\cdot\kay_2}{k_1k_2}$, but that
can be expressed in terms of the other five variables by
\begin{equation}
q^2=k_1^2+k_2^2-2 k_1 k_2\cos\theta_{12}.
\end{equation}

        Changing variables from $\{\kay_1,\kay_2\}$ to
$\{\q,k_{+},k_{-}\}$ converts \eqref{TRcont} to
\begin{eqnarray}
E^2[\Delta\Phi]
=\int_0^{k_c}dq&&\iint q^2 d^2\!\Omega_{\hat{\q}}
\int_{-q}^q dk_{-}\int_{2k_c+\abs{k_{-}}}^\infty dk_{+}
\frac{2\pi\coth\beta\frac{k_{+}+k_{-}}{4}\coth\beta\frac{k_{+}-k_{-}}{4}}
{512q}
\label{TR3D}\\
\times
\left\{
\vphantom{\abs{\int_{-T/2}^{T/2}}^2}
\right.
        &&\abs{(q^2-k_{-}^2)
        \int_{-T/2}^{T/2}dt\,\ell\Delta\Phi_\q(t)e^{ik_{-}t}
        -i4k_{-}\left[e^{i2k_{-}t}\ell\Delta\Phi_\q(t)
        \right]_{-T/2}^{T/2}}^2
\nonumber\\
        &&+\abs{(q^2-k_{-}^2)
        \int_{-T/2}^{T/2}dt\,\ell\Delta\Phi_\q(t)e^{-ik_{-}t}
        +i4k_{-}\left[e^{-i2k_{-}t}\ell\Delta\Phi_\q(t)
        \right]_{-T/2}^{T/2}}^2
\nonumber\\
        &&+\abs{(k_{+}^2-q^2)
        \int_{-T/2}^{T/2}dt\,\ell\Delta\Phi_\q(t)e^{ik_{+}t}
        +i4k_{+}\left[e^{i2k_{+}t}\ell\Delta\Phi_\q(t)
        \right]_{-T/2}^{T/2}}^2
\nonumber\\
        &&+\abs{(k_{+}^2-q^2)
        \int_{-T/2}^{T/2}dt\,\ell\Delta\Phi_\q(t)e^{-ik_{+}t}
        -i4k_{+}\left[e^{-i2k_{+}t}\ell\Delta\Phi_\q(t)
        \right]_{-T/2}^{T/2}}^2
\left.
\vphantom{\abs{\int_{-T/2}^{T/2}}^2}
\right\}.
\end{eqnarray}

\begin{figure}
\begin{center}
\epsfig{file=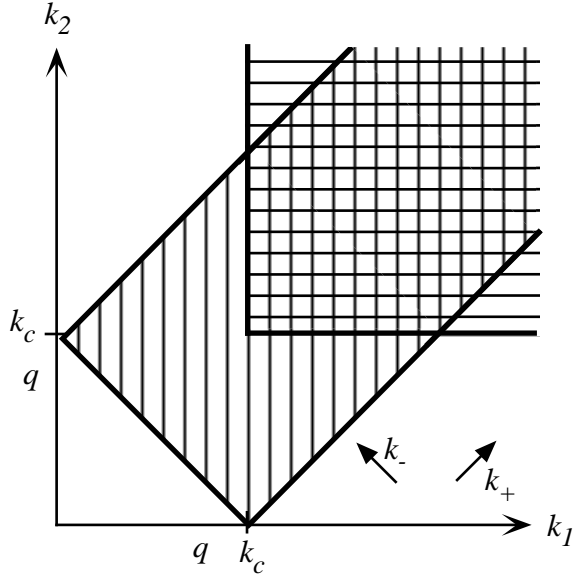}
\end{center}
\caption
{The regions of integration for \protect\eqref{TRcont}.  The inherent
geometrical restrictions $k_{+}=k_1+k_2\ge q$ and
$\abs{k_{-}}=\abs{k_1-k_2}\le q$ limit us to the region shaded
vertically, while the additional requirement that $k_1,k_2\ge k_c$
requires that the mode be in the region shaded horizontally.  Their
intersection gives the region of integration for
\protect\eqref{TRcont}.}
\label{fig:k1k2}
\end{figure}

        The Jacobian is straightforward to calculate, and the limits
of integration come from combining the restrictions $q<k_c<k_1,k_2$ on
\eqref{TRcont} with the inherent geometrical requirement that
$k_{-}^2<q^2<k_{+}^2$ (from $\abs{\cos\theta_{12}}\le 1$), as
illustrated in Fig.~\ref{fig:k1k2}.

\section{The full action}\label{sec:full}

        Returning from the modified action $S_\off$ to the full action
$S$, one finds that the result of the previous section is not
substantially changed, as demonstrated in Appendix~\ref{app:full}.

        Including the terms linear in $\phi$ to produce the action
$S_3=S_\off+\ell S_\phi$ changes the influence functional only by a
phase (as is shown by completing the square in Sec.~\ref{ssec:line}):
\begin{equation}\label{threesame}
\abs{e^{iW_3[\Phi,\Phi']}}=\abs{e^{iW_\off[\Phi,\Phi']}}.
\end{equation}

        The addition of the terms cubic in $\phi$ to restore the full
action is handled via a perturbative expansion in powers of $\ell$
Sec.~\ref{ssec:cube}.  The cubic corrections should of course be
$\mc{O}(\ell)$ or higher, but could in principle become large as
described in Sec.~\ref{ssec:breakdown} if there were enough factors of
$\beta$ in the denominator.  The calculation shows that these
corrections multiply the influence functional by a factor of order
unity:
\begin{equation}\label{eiWdiff}
e^{iW[\Phi,\Phi']}=\mc{O}(1)\times e^{iW_3[\Phi,\Phi']}.
\end{equation}
Thus we are still left with the result that
\begin{equation}\label{eiWres}
\abs{e^{iW[\Phi,\Phi']}}
\lesssim\left\{
        1+E^2[\Delta\Phi]
\right\}^{-1/4},
\end{equation}
with the suppression factor $E^2[\Delta\Phi]$ given by \eqref{TRcont}.

\subsection{A word about the perturbative analysis}

        The conclusion that
\begin{equation}
e^{iW[\Phi,\Phi']-iW_3[\Phi,\Phi']}=\mc{O}(1)
\end{equation}
is based upon an upper limit on each term in the perturbation series
(the first term is obviously unity).  There are two ways this analysis
could fail.  First, there may be cancellation among the various
$\mc{O}(1)$ terms causing the net expression to be a higher order in
$\ell$ or $\beta$.  Since this would only make $\abs{e^{iW}}$
\emph{smaller} than our estimate, it would only improve the upper limit
given by \eqref{eiWres}.

        The second is more problematic.  While each individual term is
at most $\mc{O}(1)$, the entire infinite series could be quite large,
counteracting the tendency of $e^{iW_\off}$ to become small.  This is
a shortcoming of the perturbative analysis, and there's not a lot to
be done, other than to tackle the non-perturbative
problem.\footnote{For instance, we can't use \eqref{W0} to conclude
that the $\mc{O}(1)$ factor in \eqref{eiWdiff} is unity,
since that would involve an illegal interchange of the
$\beta\rightarrow 0$ and $\ell\rightarrow 0$ limits.}  Note, however,
that we can say with confidence that $\abs{e^{iW-iW_3}}$ does not have
terms which are $\mc{O}(\ell^2/\beta^2)$, which could directly cancel
similar terms in the expansion of $\abs{e^{iW_\off}}$.  So if
$\abs{e^{iW-iW_3}}$ becomes large, it is not in the same way which
$\abs{e^{iW_3}}=\abs{e^{iW_\off}}$ becomes small.

\section{Interpretation}\label{sec:interp}

Having placed limits on the influence phase via the suppression factor
$E^2[\Delta\Phi]$ given in \eqref{TR3D}, we now consider the question
of which coarse grainings can be made to decohere for reasonable
values of the parameters $k_c$, $\beta$ and $T$.

\subsection{Which modes are suppressed?}\label{ssec:modes}

\begin{figure}
\begin{center}
\epsfig{file=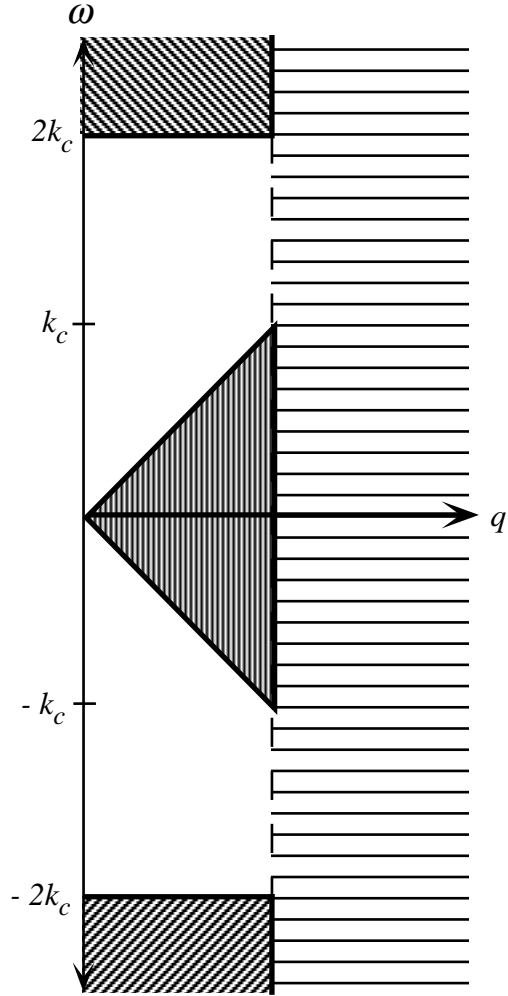,clip=,
bbllx=42,bblly=18,bburx=233,bbury=398}
\end{center}
\caption
{The modes represented in \protect\eqref{TR3D}, plotted by their
$\omega$ and $q$ values.  The modes with $q\ge k_c$ are traced over,
and so that region is shaded horizontally.  The first two terms in
\eqref{TR3D} can suppress modes with $\abs{\omega}\le q$, which are
shaded vertically, the third can suppress modes which have $\omega\ge
k_c$ and the fourth, $\omega\le -k_c$; these last two are shaded
diagonally.  Since we are concerned with coarse grainings of low
temporal frequency $\omega$ as well as spatial frequency $q$, the
first two terms are the ones of interest.}
\label{fig:modes}
\end{figure}

        Having determined that the influence functional is bounded
from above by
\begin{equation}\tag{\ref{eiWres}}
\abs{e^{iW[\Phi,\Phi']}}
\lesssim\left\{
        1+E^2[\Delta\Phi]
\right\}^{-1/4},
\end{equation}
and hence becomes small when 
\begin{eqnarray}
E^2[\Delta\Phi]
=\int_0^{k_c}&&dq\iint q^2 d^2\!\Omega_{\hat{\q}}
\int_{-q}^q dk_{-}\int_{2k_c+\abs{k_{-}}}^\infty dk_{+}
\frac{2\pi\coth\beta\frac{k_{+}+k_{-}}{4}\coth\beta\frac{k_{+}-k_{-}}{4}}
{512q}
\nonumber\\
\times
\left\{\vphantom{\abs{\int_{-T/2}^{T/2}}^2}\right.
        &&\abs{(q^2-k_{-}^2)
        \int_{-T/2}^{T/2}dt\,\ell\Delta\Phi_\q(t)e^{ik_{-}t}
        -i4k_{-}\left[e^{i2k_{-}t}\ell\Delta\Phi_\q(t)
        \right]_{-T/2}^{T/2}}^2
\nonumber\\
        &&+\abs{(q^2-k_{-}^2)
        \int_{-T/2}^{T/2}dt\,\ell\Delta\Phi_\q(t)e^{-ik_{-}t}
        +i4k_{-}\left[e^{-i2k_{-}t}\ell\Delta\Phi_\q(t)
        \right]_{-T/2}^{T/2}}^2
\nonumber\\
        &&+\abs{(k_{+}^2-q^2)
        \int_{-T/2}^{T/2}dt\,\ell\Delta\Phi_\q(t)e^{ik_{+}t}
        +i4k_{+}\left[e^{i2k_{+}t}\ell\Delta\Phi_\q(t)
        \right]_{-T/2}^{T/2}}^2
\nonumber\\
        &&\left.+\abs{(k_{+}^2-q^2)
        \int_{-T/2}^{T/2}dt\,\ell\Delta\Phi_\q(t)e^{-ik_{+}t}
        -i4k_{+}\left[e^{-i2k_{+}t}\ell\Delta\Phi_\q(t)
        \right]_{-T/2}^{T/2}}^2
\right\}\tag{\ref{TR3D}}
\end{eqnarray}
becomes large, we would like to consider when that happens.  Looking
at \eqref{TR3D}, and disregarding the surface terms (which will be
shown in Sec.~\ref{sssec:scale} to be irrelevant), we see that not all
of the space/time modes
\begin{equation}
\Delta\Phi_{\q\omega}
=\int_{-T/2}^{T/2}\frac{dt}{\sqrt{2\pi}}\Delta\Phi_\q(t)e^{i\omega t}
\end{equation}
appear.  The first two terms include only modes where
$\abs{\omega}=\abs{k_{-}}\le q$, while the last two are limited to
modes where $\abs{\omega}=k_{+}\ge 2 k_c$.  This is illustrated in
Fig.~\ref{fig:modes}.
Just as our coarse graining considers only long-wavelength modes
($q\le k_c$), it is reasonable to focus on long-period modes
($\abs{\omega}\le k_c$) as well.  Thus the limit of interest comes
from the first two terms, and we write
\begin{eqnarray}
E^2[\Delta\Phi]&\ge&
\int_0^{k_c}dq\iint q^2 d^2\!\Omega_{\hat{\q}}
\int_{-q}^q dk_{-}
\abs{(q^2-k_{-}^2)
\sqrt{2\pi}\ell\Delta\Phi_{\q k_{-}}
-i4k_{-}\left[e^{i2k_{-}t}\ell\Delta\Phi_\q(t)
\right]_{-T/2}^{T/2}}^2
\nonumber\\
&&
\times
\int_{2k_c+\abs{k_{-}}}^\infty dk_{+}
\frac{2\pi\coth\beta\frac{k_{+}+k_{-}}{4}\coth\beta\frac{k_{+}-k_{-}}{4}}
{256q}.
\end{eqnarray}
        The factor
\begin{equation}
R=\int_{2k_c+\abs{k_{-}}}^\infty dk_{+}
\frac{2\pi\coth\frac{\beta k_1}{2}\coth\frac{\beta k_2}{2}}
{256q}
\end{equation}
can be evaluated, to leading order in $\beta$, by noting that
\begin{equation}
\coth\eta_1\coth\eta_2=
\frac{\cosh\eta_1\cosh\eta_2}{\sinh\eta_1\sinh\eta_2}
=\frac{\cosh\eta_{+}+\cosh\eta_{-}}{\cosh\eta_{+}-\cosh\eta_{-}}
=1+\frac{2\cosh\eta_{-}}{\cosh\eta_{+}-\cosh\eta_{-}},
\end{equation}
so that
\begin{equation}
R=
\int_{2k_c+\abs{k_{-}}}^\infty \frac{2\pi\,dk_{+}}
{256q}
\left(
        1+\frac{2\cosh\frac{\beta k_{-}}{2}}
        {\cosh\frac{\beta k_{+}}{2}-\cosh\frac{\beta k_{-}}{2}}
\right)
=R_0+\int_{2k_c+\abs{k_{-}}}^\infty \frac{2\pi\,dk_{+}}{256q}.
\end{equation}
Now,
\begin{equation}\label{R0}
R_0=
\int_{2k_c+\abs{k_{-}}}^\infty \frac{2\pi\,dk_{+}}{256q}
\frac{2\cosh\frac{\beta k_{-}}{2}}
{\cosh\frac{\beta k_{+}}{2}-\cosh\frac{\beta k_{-}}{2}}
=\frac{4(2\pi)\coth\frac{\beta\abs{k_{-}}}{2}}
{256q\beta}
\ln\left(
        \frac{\sinh\beta\frac{k_c+\abs{k_{-}}}{2}}
        {\sinh\frac{\beta k_c}{2}e^{\beta\abs{k_{-}}/2}}
\right).
\end{equation}
Again, since we only expect a useful answer when small $\beta$ causes
perturbation theory to break down, we look at the leading terms in
$\beta$, working in the high-temperature limit $\beta k_c\gg 1$.  (See
Sec.~\ref{sssec:scale} for the physical significance of this.)  In this
limit, \eqref{R0} becomes
\begin{equation}
R_0
=\frac{8(2\pi)}
{256q\beta^2\abs{k_{-}}}
\ln\left(1+\frac{\abs{k_{-}}}{k_c}\right);
\end{equation}
since $R-R_0$ is independent of $\beta$, the leading term in $R$
is\footnote{Of course, this is a dubious approximation, since $R-R_0$,
while down by a factor of $\beta^2$ from $R_0$, is ultraviolet
divergent.  However, any suitable well-behaved regulation of the
result will give a result which agrees with $R_0$ to
$\mc{O}(\beta^{-2})$ when the $\beta\rightarrow 0$ limit is taken
before the cutoff limit.  Note also that our perturbative analysis has
ignored terms like $\ell^2 (R-R_0)$, which are perturbatively small in
$\ell$ without having corresponding factors of $\beta$.  One might
hope that such terms will cancel the divergence in $R-R_0$.  However,
this turns out not to be the case, as can be seen by calculating all
of the $\mc{O}(\ell^2)$ terms in $\abs{e^{iW_\off}}$.}
\begin{equation}
R=\frac{2\pi}{32q\beta^2\abs{k_{-}}}
\ln\left(1+\frac{\abs{k_{-}}}{k_c}\right),
\end{equation}
so
\begin{equation}
E^2[\Delta\Phi]
\gtrsim\int_0^{k_c}dq\iint q^2 d^2\!\Omega_{\hat{\q}}
\int_{-q}^q \frac{2\pi\,d\omega}{32q\beta^2\abs{\omega}}
\ln\left(1+\frac{\abs{\omega}}{k_c}\right)
\abs{(q^2-\omega^2)\sqrt{2\pi}\ell\Delta\Phi_{\q\omega}
-i4\omega \left[e^{i2\omega t}\ell\Delta\Phi_\q(t)
\right]_{-T/2}^{T/2}}^2.\label{TRphys}
\end{equation}

\subsection{Practical coarse grainings}\label{ssec:CG}

\subsubsection{The physical scales}\label{sssec:scale}

        The expression \eqref{TRphys} has three parameters, $k_c$,
$\beta$, and $T$, which are not integrated over.  The scale $k_c$ for
division into SWMs and LWMs can be tailored to the coarse graining to
give the strongest possible results, while the other two are features
of the model.  As alluded to in Sec.~\ref{sec:GRscalar}, the time scale
$T$ over which we expect the Minkowski space model to be valid should
be slightly below the Hubble scale $H_0^{-1}$.  In suitable units,
this gives
\begin{equation}
T\lesssim H_0^{-1}\sim 10^{10}\,\textrm{yr}\sim 10^{29}\,\textrm{cm}.
\end{equation}
This is so large that it allows us to set $T$ much larger than all the
other scales in the problem.  In particular, it means that the cross
terms in
\begin{eqnarray}
\left|\vphantom{]_{-T/2}^{T/2}}\right.
(q^2-\omega^2)\sqrt{2\pi}&&\ell\Delta\Phi_{\q\omega}
-i4\omega \left.\left[e^{i2\omega t}\ell\Delta\Phi_\q(t)
\right]_{-T/2}^{T/2}\right|^2
\\
=&&2\pi\abs{(q^2-\omega^2)\ell\Delta\Phi_{\q\omega}}^2
+i\sqrt{2\pi}(q^2-\omega^2)\ell\Delta\Phi_{\q\omega}
4\omega\left(e^{-i\omega T}\ell{\Delta\Phi_{\q f}}^*
             -e^{i\omega T}\ell{\Delta\Phi_{\q i}}^*\right)
\nonumber\\
\nonumber
&&-i\sqrt{2\pi}(q^2-\omega^2)\ell\Delta\Phi_{\q\omega}^*
4\omega\left(e^{i\omega T}\ell\Delta\Phi_{\q f}
             -e^{-i\omega T}\ell\Delta\Phi_{\q i}\right)
+16\omega^2\abs{
 e^{i\omega T}\ell\Delta\Phi_{\q f}-e^{-i\omega T}\ell\Delta\Phi_{\q i}}^2
\end{eqnarray}
will oscillate rapidly and vanish when $\omega$ is integrated over,
leaving
\begin{equation}
2\pi\abs{(q^2-\omega^2)\ell\Delta\Phi_{\q\omega}}^2
+
16\omega^2\abs{
 e^{i\omega T}\ell\Delta\Phi_{\q f}-e^{-i\omega T}\ell\Delta\Phi_{\q i}}^2
\ge 2\pi\abs{(q^2-\omega^2)\ell\Delta\Phi_{\q\omega}},
\end{equation}
so
\begin{equation}
E^2[\Delta\Phi]
\gtrsim\int_0^{k_c}dq\iint q^2 d^2\!\Omega_{\hat{\q}}
\int_{-q}^q \frac{2\pi\,d\omega}{32q\beta^2\abs{\omega}}
\tag{\ref{TRphys}$'$}\label{TRsimp}
\ln\left(1+\frac{\abs{\omega}}{k_c}\right)
\abs{(q^2-\omega^2)\sqrt{2\pi}\ell\Delta\Phi_{\q\omega}}^2.
\end{equation}

        Turning our attention to the inverse temperature $\beta$, we
might reasonably treat the high-temperature thermal state $\rho_\phi$
as corresponding to the cosmic graviton background
radiation \cite{K&T}, which has a temperature on the order of
$1\,\textrm{K}$.  This means that in suitable units,
\begin{equation}
\beta\sim\frac{1}{1\,\textrm{K}}\sim\frac{1}{10^{-4}\,\textrm{eV}}
\sim 10^{-1}\,\textrm{cm}.
\end{equation}
This is the most severe limit to the usefulness of the calculations in
this work.  It means that to be in the high-temperature limit $\beta
k_c\ll 1$, we need to have the cutoff scale $k_c^{-1}$ dividing
``short'' and ``long'' wavelengths be above the millimeter scale.
While we don't expect to have laboratory data on millimeter-scale
oscillations of vacuum gravity any time soon (contrast this scale to
the length corresponding to a typical component of the curvature
tensor at the surface of a $1M_{\odot}$ black hole, which is
$GM_{\odot}\sim 1\,\textrm{km}\sim 10^6\beta$), it might be a bit
surprising to learn that coarse grainings corresponding to
micron-scale variations in the gravitational field do not decohere.
At any rate, that is \emph{not} the prediction of this work, even
assuming that the analysis of the conformally flat toy model is an
accurate indicator of the behavior of the full theory.  First, a
perturbative analysis of vacuum gravity simply cannot make fruitful
predictions outside of the perturbative regime.  It is quite possible
that for lower temperatures, non-perturbative effects can cause the
influence functional to become small for large $\Delta\Phi$.  And of
course, this analysis only models the decoherence of the vacuum
gravitational field induced by gravity itself.  If the gravitational
field is coupled to some form of matter, unobserved modes of the
matter can also induce decoherence, as described in
\cite{JJH89,padman}.  So this work suggests an encouraging lower limit
on the effectiveness of the decoherence of spacetime itself without
the assistance of additional matter fields.

\subsubsection{Field averages}

        When constructing the sum-over-histories generalized quantum
mechanics of a field theory, it is useful to coarse grain by values of
a field average over a particular region\cite{leshouches}. By defining
the average
\begin{equation}\label{avedef}
\langle\ell\varphi\rangle
=\int d^3\!x \int_{-T/2}^{T/2} dt\, w(\x,t)\ell\varphi(\x,t)
\end{equation}
with a weighting function $w(\x,t)$ it is possible to study the
behavior of different Fourier modes of the field via the choice of
$w(\x,t)$.  For now we take the weighting function to be normalized,
\begin{equation}\label{wnorm}
\int d^3\!x \int_{-T/2}^{T/2} dt\, w(\x,t)=1,
\end{equation}
but later we will relax that restriction to allow for averages in
Fourier space which do not include the zero mode.

        As an example of a field average, recall the connection of our
scalar field theory to the theory of a conformally flat metric
discussed in Sec.~\ref{sec:GRscalar}.  As described there, the
fractional deviation of the volume of a spacetime region $S$ from its
volume in flatthe background metric, $\int
d^{\Ddim+1}\!x\sqrt{\abs{g}}\,w_S(x)-1$, is a field average (given by
\eqref{volnorm}) which in the $D=3$ case currently being considered is
\begin{equation}
\frac{2}{3}(2\pi)^{3/2}\langle\ell\varphi\rangle_S,
\end{equation}
with the weighting function taken to be the characteristic function
for $S$
\begin{equation}\label{wS}
w_S(x)=
\cases{
V_0^{-1} & $x\in S$\cr
0 & $x\notin S$.
}
\end{equation}
($V_0$ is of course the background volume of $S$.)

        In terms of Fourier modes, a general field average becomes
\begin{equation}
\langle\ell\varphi\rangle
=\int d^3\!q\, d\omega\, w_{\q\omega}^*\ell\varphi_{\q\omega},
\end{equation}
where we have approximated the sum over $\omega$ values separated by
$\delta\omega=2\pi/T$ by an integral, and assumed that $w(\x,t)$
vanishes as $t\rightarrow\pm T/2$, so that it is acceptable to replace
the field $\varphi_{\q}(t)$ by its periodic counterpart
\begin{equation}
\varphi_{\q}^{\textsc{p}}(t)=\int \frac{d\omega}{\sqrt{2\pi}}
\,\varphi_{\q\omega}e^{-i\omega t}
=\cases{
\frac{1}{2}(\varphi_{\q f}+\varphi_{\q i}) & $t=\pm\frac{T}{2}$\cr
\varphi_{\q}(t) & $\!\!\!\!\!\!\!-\frac{T}{2}<t<\frac{T}{2}$.
}
\end{equation}
If $w_{\q}$ contains only modes with $\q\in\mc{L}$
[i.e. $w_{\q}=\Theta(k_c-q)w_{\q}$], then we can write this average as
\begin{equation}
\langle\ell\Phi\rangle
=\int\limits_{\mc{L}} d^3\!q\, d\omega\,
w_{\q\omega}^*\ell\Phi_{\q\omega}.
\end{equation}
The normalization condition \eqref{wnorm} becomes
$w_{{\bf 0}0}=(2\pi)^{-2}$, so a useful field average might be
\begin{equation}
\frac{2}{3}(2\pi)^{3/2}\langle\ell\Phi\rangle
=\frac{2}{3}\int_0^{\Delta q/2}dq\iint q^2 d^2\!\Omega_{\hat{\q}}
\int_{-\Delta\omega/2}^{\Delta\omega/2}\frac{d\omega}{\sqrt{2\pi}}
\Phi_{\q\omega},
\end{equation}
where the width of the smoothing function in Fourier space is
\begin{mathletters}
\begin{eqnarray}
\Delta q&\sim&\frac{1}{\Delta x}\\
\Delta\omega&\sim&\frac{1}{\Delta t}
\end{eqnarray}
\end{mathletters}
and the origin of the spatial {\coord}s has been chosen to correspond
with the center of $w(\x,t)$.  To consider a group of modes not
centered about the constant mode, we shift the center of the group of
Fourier modes by $\q_0$ and $\omega_0$, while keeping the mode volume
the same, giving another dimensionless quantity
\begin{equation}\label{CG}
\frac{2}{3}(2\pi)^{3/2}\wt{\langle\ell\Phi\rangle}
=
\frac{2}{3}\int_{q_0-\Delta q/2}^{q_0+\Delta q/2}dq
\iOint q^2 d^2\!\Omega_{\hat{\q}}
\int_{\omega_0-\Delta\omega/2}^{\omega_0+\Delta\omega/2}
\frac{d\omega}{\sqrt{2\pi}}(\Phi_{\q\omega}+\Phi_{-\q,-\omega}),
\end{equation}
where the solid angle integrated over is centered about $\hat{\q}_0$
and is chosen to preserve the mode volume:
\begin{equation}
\frac{4\pi(\Delta q/2)^3}{3}
=2\Omega\int_{q_0-\Delta q/2}^{q_0+\Delta q/2}q^2 dq
=2\Omega\frac{(q_0+\Delta q/2)^3-(q_0-\Delta q/2)^3}{3},
\end{equation}
so
\begin{equation}
\Omega=\frac{\pi(\Delta q)^2}{12q_0^2 + (\Delta q)^2}.
\end{equation}

\subsubsection{The influence phase}

Now we can cast \eqref{TRphys} into a useful form, so long as
$q_0-\Delta q/2\ge\abs{\omega_0}+\Delta\omega$:
\begin{eqnarray}
E^2[\Delta\Phi]
&\gtrsim&\int_0^{k_c}dq\iint q^2 d^2\!\Omega_{\hat{\q}}
\int_{-q}^q \frac{2\pi d\omega}{32q\beta^2\abs{\omega}}
\ln\left(1+\frac{\abs{\omega}}{k_c}\right)
\abs{(q^2-\omega^2)\sqrt{2\pi}\ell\Delta\Phi_{\q\omega}}^2
\nonumber\\
&\ge&\int_{q_0-\Delta q/2}^{q_0+\Delta q/2}dq
\iOint q^2 d^2\!\Omega_{\hat{\q}}
\int_{\omega_0-\Delta\omega/2}^{\omega_0+\Delta\omega/2}d\omega\,
\frac{(2\pi)^2(q^2-\omega^2)^2}{32q\beta^2\abs{\omega}}
\ln\left(1+\frac{\abs{\omega}}{k_c}\right)
\left(
   \abs{\ell\Delta\Phi_{\q\omega}}^2+\abs{\ell\Delta\Phi_{-\q,-\omega}}^2
\right)\label{TRuse}
\end{eqnarray}
The strongest result will be obtained if we take $k_c=q_0+\Delta q/2$.
If $\Delta\omega$ and $\Delta q$ are small relative to $\omega_0$ and
$q_0$ (which means large $\Delta t$ and $\Delta x$), we can
approximate
\begin{eqnarray}
E^2[\Delta\Phi]
&\gtrsim&\Theta(q_0-\abs{\omega_0})
\frac{(2\pi)^2(q_0^2-\omega_0^2)^2}{32q_0\beta^2\abs{\omega_0}}
\ln\left(1+\frac{\abs{\omega_0}}{q_0}\right)
\int_{q_0-\Delta q/2}^{q_0+\Delta q/2}dq
\iOint q^2 d^2\!\Omega_{\hat{\q}}
\int_{\omega_0-\Delta\omega/2}^{\omega_0+\Delta\omega/2}d\omega\,
\left(
   \abs{\ell\Delta\Phi_{\q\omega}}^2+\abs{\ell\Delta\Phi_{-\q,-\omega}}^2
\right)
\nonumber\\
&\approx&\Theta(q_0-\abs{\omega_0})
\frac{(2\pi)^2(q_0^2-\omega_0^2)^2}{32q_0\beta^2\abs{\omega_0}}
\ln\left(1+\frac{\abs{\omega_0}}{q_0}\right)
\frac{(2\pi)^2\abs{\wt{\langle\ell\Delta\Phi\rangle}}^2}
        {\pi\Delta\omega(\Delta q)^3/6}
\end{eqnarray}
so that the influence phase is bounded by
\begin{equation}
\abs{e^{iW[\Phi,\Phi']}}
\label{inflmode}
\lesssim\left\{
1
+\frac{3\pi^4(q_0^2-\omega_0^2)^2
          \abs{\wt{\langle\ell\Delta\Phi\rangle}}^2}
        {q_0\beta^2\abs{\omega_0}\Delta\omega(\Delta q)^3}
        \ln\left(1+\frac{\abs{\omega_0}}{q_0}\right)
\right\}^{-1/4}.
\end{equation}
This means that if
\begin{equation}
\frac{3\pi^4(q_0^2-\omega_0^2)^2
          \abs{\wt{\langle\ell\Delta\Phi\rangle}}^2}
        {q_0\beta^2\abs{\omega_0}\Delta\omega(\Delta q)^3}
        \ln\left(1+\frac{\abs{\omega_0}}{q_0}\right)
\gg 1,
\end{equation}
the decoherence functional $D[\Phi,\Phi']$ corresponding to
$\wt{\langle\ell\Phi\rangle}$ and $\wt{\langle\ell\Phi'\rangle}$
separated by $\wt{\langle\ell\Delta\Phi\rangle}$ will be small.  This
limit corresponds to
\begin{equation}
\frac{2}{3}(2\pi)^{3/2}\abs{\wt{\langle\ell\Delta\Phi\rangle}}
\gg\frac{4\beta\sqrt{2q_0\abs{\omega_0}\Delta\omega(\Delta q)^3}}
{3(q_0^2-\omega_0^2)\sqrt{3\pi}}
\left[\ln\left(1+\frac{\abs{\omega_0}}{q_0}\right)\right]^{-1/2};
\label{lim}
\end{equation}
Considering the static limit $\abs{\omega_0}\ll q_0$ for simplicity,
\eqref{lim} becomes
\begin{equation}
\frac{2}{3}(2\pi)^{3/2}\abs{\wt{\langle\ell\Delta\Phi\rangle}}
\gg\frac{4\beta\sqrt{2q_0\abs{\omega_0}\Delta\omega(\Delta q)^3}}
{3\pi^2 q_0^2\sqrt{3\pi}}
\left[\frac{\abs{\omega_0}}{q_0}\right]^{-1/2}
\label{limfin}
=\frac{4\beta q_0}{3}
\sqrt{\frac{2\Delta\omega(\Delta q)^3}{3\pi q_0^4}}.
\end{equation}
For sufficiently small $\Delta\omega$ and $\Delta q$ (which
corresponds to averaging over a large spacetime region), the right
hand side of \eqref{limfin} becomes small, and thus \eqref{limfin} can
hold even when the quantity
$\frac{2}{3}(2\pi)^{3/2}\abs{\wt{\langle\ell\Delta\Phi\rangle}}$
representing the perturbation due to the metric is small, justifying
the use of perturbation theory.

\begin{figure}
\begin{center}
\epsfig{file=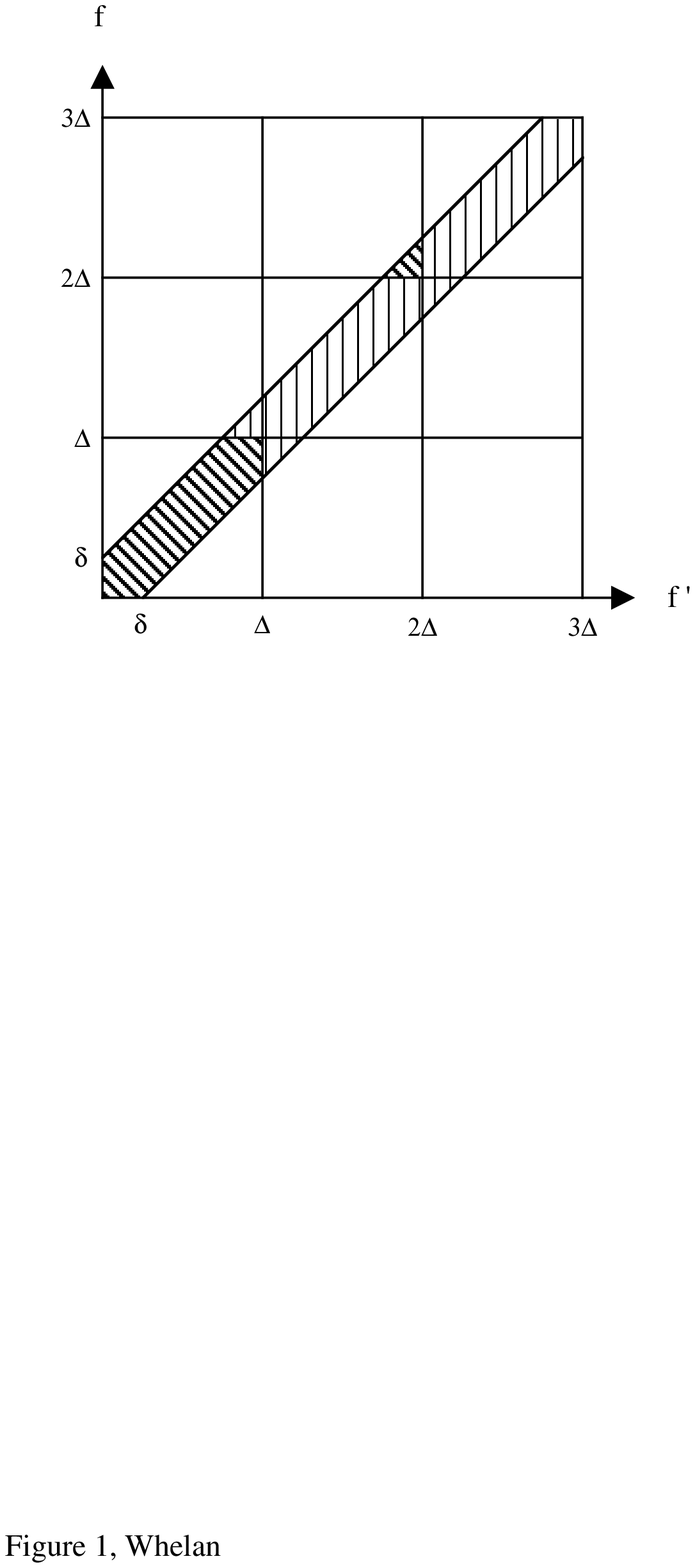,height=3.5in,width=3.5in,clip=,%
bbllx=80,bblly=450,bburx=373,bbury=743}
\end{center}
\caption{The suppression of
$D(n,n')=\int_{n\Delta}^{(n+1)\Delta}df\int_{n'\Delta}^{(n'+1)\Delta}df'
G(f,f')$ via the influence functional $e^{iW}$.  If $G(f-f')$ is
suppressed for $\abs{f-f'}\protect\gtrsim\delta$, integrals of $G(y)$
over regions two or more spots off the diagonal ($\abs{n-n'}\ge 2$)
will be negligible.  Squares on the diagonal ($n=n'$) have a region of
area $2\Delta\delta-\delta^2$ over which $G(f,f')$ is appreciable.
Squares one spot off the diagonal ($\abs{n-n'}=1$) include some
non-negligible values of $G(f,f')$, but only in a triangular region of
area $\delta^2/2$.  Thus $D(n,n\pm 1)$ should be suppressed by a
factor of $\delta/\Delta$ relative to $D(n,n)$.  Compare Fig.~1 of
\protect\cite{classeq}.}
\label{fig:dropoff}
\end{figure}

        So a coarse graining which should decohere is one consisting
of a set of alternatives $\{c_n\}$ which correspond to
$\frac{2}{3}(2\pi)^{3/2}\wt{\langle\ell\Phi\rangle}\in\Delta_n
=[n\Delta,(n+1)\Delta)$.  The decoherence functional for such a coarse
graining will be
\begin{equation}
D(n,n')
=\!\!\!\int\limits_{\wt{\langle\ell\Phi\rangle}\in\Delta_n}
\!\!\!\!\D\Phi\!\!\!
\int\limits_{\wt{\langle\ell\Phi'\rangle}\in\Delta_{n'}}\!\!\!\!
\D\Phi'D[\Phi,\Phi']
\label{CGDF}
=\int_{n\Delta}^{(n+1)\Delta}df\int_{n'\Delta}^{(n'+1)\Delta}df'
G(f,f'),
\end{equation}
where
\addtocounter{equation}{-1}
\begin{mathletters}
\begin{equation}
G(f,f')
=\int\D\Phi\,\D\Phi' 
\rho_\Phi(\Phi_i,\Phi'_i)\delta(\Phi'_f-\Phi_f)
e^{i(S_\Phi[\Phi]-S_\Phi[\Phi']+W[\Phi,\Phi'])}
\delta\!\left(f-\frac{2}{3}(2\pi)^{3/2}\wt{\langle\ell\Phi\rangle}\right)
\delta\!\left(f'-\frac{2}{3}(2\pi)^{3/2}\wt{\langle\ell\Phi'\rangle}\right)
\end{equation}
\end{mathletters}
Equation~\eqref{inflmode} shows that $G(f,f')$ should be suppressed
by the influence functional when
\begin{equation}
f-f'\gtrsim\delta=\frac{4\beta q_0}{3}
\sqrt{\frac{2\Delta\omega(\Delta q)^3}{3\pi q_0^4}}.
\end{equation}
As long as the size $\Delta$ of the bins is much larger than $\delta$,
the off-diagonal elements of $D(n,n')$ with $\abs{n-n'}\ge 2$ will
involve integrals only over the suppressed region, while the elements
with $\abs{n-n'}=1$ should be down from the diagonal elements by a
factor of $\delta/\Delta$.  (Fig.~\ref{fig:dropoff})

        To express the result in terms of a familiar measure of
decoherence, we can define a \emph{decoherence time}
$T_{\text{dec}}=2\pi/\Delta\omega$, which is the temporal extent of a
weighting function leading to a decohering coarse graining.  Solving
for $T_{\text{dec}}$ gives
\begin{equation}
T_{\text{dec}}\sim\frac{64}{27}\frac{\beta^2(\Delta q)^3}
{\delta^2q_0^2}
\sim\frac{512\pi^3}{27}\frac{\beta^2}{\varepsilon^2\Delta^2q_0^2V},
\end{equation}
where $V=(2\pi/\Delta q)^3$ is the spatial volume over which the
weighting function is non-negligible, $q_0$ is the spatial frequency
at which it oscillates, $\Delta$ is the size of the bins in our coarse
graining, and $\varepsilon$ is our standard for approximate
decoherence (how small the off-diagonal elements of the decoherence
functional must be).

\subsection{Impractical coarse grainings}\label{ssec:imprac}

\begin{figure}
\begin{center}
\epsfig{file=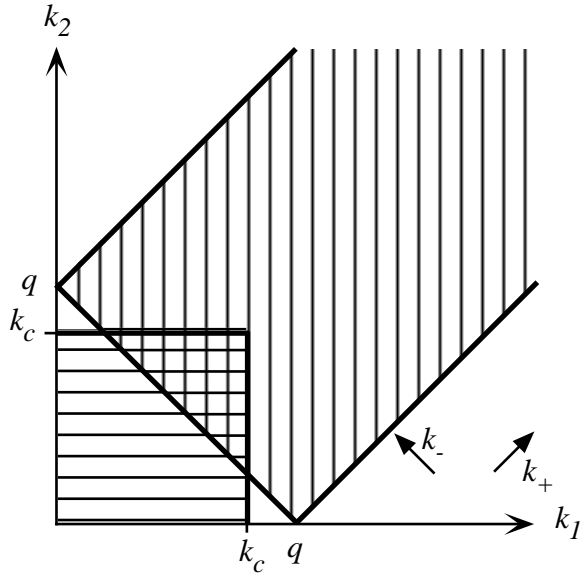}
\end{center}
\caption{The modified regions of integration for
\protect\eqref{TRcont} when the system is made up of short-wavelength
modes and the environment of long-wavelength modes.  As in
Fig.~\protect\ref{fig:k1k2}, the geometrical restrictions
$k_{+}=k_1+k_2\ge q$ and $\abs{k_{-}}=\abs{k_1-k_2}\le q$ limit us to
the region shaded vertically.  Now the requirement producing the
horizontally-shaded region is $k_1,k_2\ge k_c$.  Note that in contrast
with the regions of integration in Fig.~\protect\ref{fig:k1k2}, the
ranges of both the $k_{+}$ and $k_{-}$ integrations are finite.}
\label{fig:newk1k2}
\end{figure}

\begin{figure}
\begin{center}
\epsfig{file=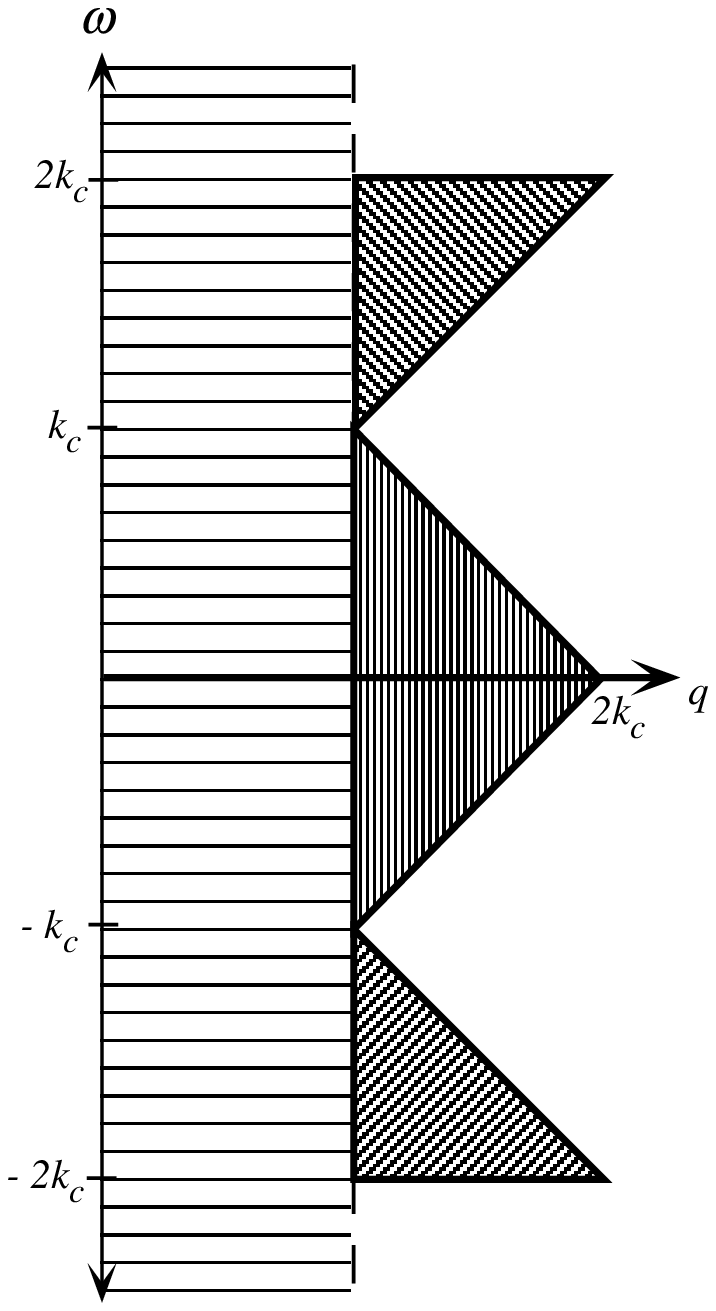,clip=,
bbllx=42,bblly=18,bburx=246,bbury=398}
\end{center}
\caption
{The modes represented in the suppression term
$E^2[\Delta\Phi]$, plotted by their $\omega$ and $q$
values.  The modes with $q\le k_c$ are traced over, and so that region
is shaded horizontally.  The terms with $\omega=\pm k_{-}$ can
suppress modes with $\abs{\omega}\le 2k_c-k_{+}$, which are shaded
vertically, those with $\omega=k_{+}$ can suppress modes which have
$q\le\omega\le2k_c$ and those with $\omega=-k_{+}$ can suppress
$-q\ge\omega\ge-2k_c$; these last two are shaded diagonally.}
\label{fig:newmodes}
\end{figure}

        A question one might like to ask is whether the decoherence
exhibited in the previous sections relied upon the fact that the modes
of interest were the long-wavelength ones, or if the mere fact that
\emph{some} sufficiently large group of modes is traced over is enough
to produce decoherence.  In this section, we show that we obtain a
similar result if the identification of system and environment in
\eqref{Phiphi} are now reversed:
\begin{mathletters}
\begin{eqnarray}
\Phi_\N(t)&=&\varphi_\N(t)\Theta(q-k_c) \\
\phi_\M(t)&=&\varphi_\M(t)\Theta(k_c-q),
\end{eqnarray}
\end{mathletters}
so that now $\N\in\mc{S}$ and $\M\in\mc{L}$.  Most of the calculation
carries through unchanged until it comes to determining the limits of
integration in Sec.~\ref{sssec:3D}.  There the geometrical limits are
still $k_{-}^2<q^2<k_{+}^2$, but now we have $k_1,k_2<k_c<q$, which
gives the region of integration shown in Fig.~\ref{fig:newk1k2}.
The limits on the integration variables in \eqref{TR3D} are thus
changed so that $q$ runs from $k_c$ to $2k_c$, $k_{+}$ from $q$ to
$2k_c$, and $k_{-}$ from $-(2k_c-k_{+})$ to $2k_c-k_{+}$ (or,
equivalently, $k_{-}$ runs from $-(2k_c-q)$ to $2k_c-q$ and $k_{+}$
from $q$ to $2k_c-\abs{k_{-}}$, with the same limits on the $q$
integration).  Moving to Sec.~\ref{ssec:modes}, we find the new
regions of potentially suppressed frequencies, illustrated in
Fig.~\ref{fig:newmodes}.
The terms in \eqref{TR3D} with $\abs{\omega}=\abs{k_{-}}$ will have
$\abs{\omega}\le 2k_c-q$, and are shaded vertically in
Fig.~\ref{fig:newmodes}, while those with $\abs{\omega}=k_{+}$ will
have $\abs{\omega}$ between $q$ and $2k_{c}$ and are shaded
diagonally.

        Since $k_{-}\le k_c\le q\le k_{+}\le 2k_{c}$ we can express
the suppression factor in the limit that $2\beta k_{c}\ll 1$ as
\begin{eqnarray}
E^2[\Delta\Phi]
&\gtrsim&\int_{k_c}^{2 k_c}dq\iint q^2 d^2\!\Omega_{\hat{\q}}
\frac{2\pi}{64q\beta^2}
\left[
        2\int_{-(2k_c-q)}^{2k_c-q}\frac{d\omega}{\abs{\omega}}
        \ln\left(\frac{k_c-\abs{\omega}}{q-\abs{\omega}}
                 \,\frac{q+\abs{\omega}}{k_c}\right)
\right.
\nonumber\\
&&\phantom{\int_{k_c}^{2 k_c}dq\iint q^2 d^2\!\Omega_{\hat{\q}}}
        \times
        \abs{(q^2-\omega^2)\sqrt{2\pi}\ell\Delta\Phi_{\q\omega}
        -i4\omega \left[e^{i2\omega t}\ell\Delta\Phi_\q(t)
        \right]_{-T/2}^{T/2}}^2
\nonumber\\
        &&+\int_{q}^{2k_c}\frac{d\omega}{\omega}
        \ln\left(\frac{\omega}{k_c}-1\right)
        \left(
                \abs{(\omega^2-q^2)\sqrt{2\pi}\ell\Delta\Phi_{\q\omega}
                +i4\omega \left[e^{i2\omega t}\ell\Delta\Phi_\q(t)
                \right]_{-T/2}^{T/2}}^2
        \right.
\nonumber\\
&&\phantom{+\int_{q}^{2k_c}\frac{d\omega}{\omega}
\ln\left(\frac{\omega}{k_c}-1\right)\Bigl(}
\left.
        \left.
                +\abs{(\omega^2-q^2)\sqrt{2\pi}\ell\Delta\Phi_{\q,-\omega}
                -i4\omega \left[e^{-i2\omega t}\ell\Delta\Phi_\q(t)
                \right]_{-T/2}^{T/2}}^2
        \right)
\right].
\end{eqnarray}
        Focussing on the modes shaded horizontally in
Fig.~\ref{fig:newmodes} and neglecting the boundary terms as described
in Sec.~\ref{sssec:scale}, we see that the lower limit on the
suppression factor becomes
\begin{equation}
E^2[\Delta\Phi]
\gtrsim\int_{k_c}^{2k_c}dq\iint q^2 d^2\!\Omega_{\hat{\q}}
\int_{-(2k_c-q)}^{2k_c-q} \frac{2\pi d\omega}{32q\beta^2\abs{\omega}}
\ln\left(1+\frac{\abs{\omega}}{k_c}\right)
\abs{(q^2-\omega^2)\sqrt{2\pi}\ell\Delta\Phi_{\q\omega}}^2,
\end{equation}
which is, other than the limits of integration on $q$ and $\omega$,
the same as that given in \eqref{TRuse}.  Thus we conclude that, at
least for these groups of modes with $q_0>\abs{\omega_0}$, \emph{the
tendency of unobserved modes to induce decoherence is just as
effective whether they are of shorter or longer wavelength.}

        There is some precedent for this result in, for example,
\cite{JJH89}, where the \emph{long}-wavelength modes of an additional
scalar field induced decoherence in the gravitational field.

\section{Conclusions}

        This work has demonstrated that, in a scalar field theory
obtained by perturbative expansion of the Nordstr\"{o}m-Einstein-Fokker
action (which is given by the Einstein-Hilbert action restricted to
conformally flat metrics), some coarse grainings which restrict only
the long wavelength modes of the field should decohere.  Using the
self-interaction of this theory\footnote{In terms of an appropriate
set of variables; in any number of dimensions, there is a
reparametrization of the theory which is non-interacting, but it is
the self-interaction in terms of the variables defining the
coarse-graining which is relevant for decoherence; see
footnote~\ref{fn:free}.
}, which has a form
analogous to that obtained by a perturbative expansion of GR, the
ignored short-wavelength degrees of freedom can destroy quantum
coherence between different long-wavelength alternatives.  This lack
of quantum-mechanical interference is a prerequisite for classical
behavior of spacetime on large scales.  The present result
demonstrates that in some cases the gravitational self-interaction, as
represented by this toy model, is sufficient to induce decoherence
without adding any matter fields.

        As demonstrated in Sec~\ref{ssec:imprac}, the central feature
of this mechanism is the division into a system and an environment.
The split can also induce decoherence when we coarse grain by the
short-wavelength features and let the long-wavelength modes act as an
environment.

        The decoherence properties were studied by calculating the
influence functional $e^{iW}$ between pairs of long-wavelength
histories, which describes the effect of tracing out the
short-wavelength modes.  Decoherence is expected when its absolute
value $\abs{e^{iW[\Phi,\Phi']}}$ becomes small for sufficiently large
differences between the long-wavelength configurations $\Phi$ and
$\Phi'$.

        Even though the influence functional is unity to lowest order
in the coupling constant $\ell$, and one might normally assume that
perturbative corrections cannot make $\abs{e^{iW}}$ much smaller than
one, decoherence is still possible if there is a second small
parameter.  In our case, this was accomplished by working in the
high-temperature regime where the inverse temperature $\beta$ of the
thermal state describing the SWMs was small.  Then, as described in
Sec.~\ref{ssec:breakdown}, terms which were higher order in the
coupling $\ell$ could still become large for high temperature if they
were proportional to, for example, $(\ell/\beta)^2$.  This made it
possible to find $\abs{e^{iW}}\ll 1$, while still allowing us to treat
terms higher order in $\ell$ as small if they did not have
corresponding powers of $\beta^{-1}$.

        The $\ell/\beta$ terms in the influence functional were
handled non-perturbatively for the terms in the action which are
quadratic or linear in the SWMs, but the cubic terms in the action
were analyzed using a perturbative expansion.  That expansion showed
that while there are corrections which go like
$\frac{\mc{O}(\ell)}{\mc{O}(\ell)+\mc{O}(\beta)}$ or
$\frac{\mc{O}(\beta)}{\mc{O}(\ell)+\mc{O}(\beta)}$, those are at
largest $\mc{O}(1)$, and there are no $\mc{O}(\ell/\beta)$ terms to
cancel out the effect from the quadratic action.

        The reliance on perturbative analysis is one of the
limitations of this result.  It means that we can only analyze the
question of decoherence in the high-temperature limit, defined by
$\beta k_c\ll 1$, where $k_c$ is the momentum which divides SWMs from
LWMs.  If the temperature of the SWM thermal state is taken to be that
of the present-day cosmic graviton background, the length scale
corresponding to this limit is on the order of a millimeter.

        Another problem comes from the non-renormalizability of our
derivative action (a property it shares with GR itself).  While the
terms in the influence functional proportional to $(\ell/\beta)^2$ are
finite, there are terms proportional to $\ell^2$ alone which are
ultraviolet divergent.  We were able to ignore those by working in the
high-temperature limit, but they may provide another way in which
perturbation theory breaks down, demanding a fully non-perturbative
analysis.

        Before moving to a possible non-perturbative analysis, perhaps
using the Regge calculus \cite{regge} to skeletonize geometry, another
improvement of this work would be to see if the scalar field result is
modified by considering a model where the full tensor nature of the
theory is exhibited.  In a full quantum theory of gravity, it is not
the conformal variations of the metric which are expected to be the
dynamical degrees of freedom.  However, since the interaction should
have the same form, we may expect that features of the present result
will survive.  In particular, the quantity $k_c\beta$ is likely to be
important.

        And finally, the focus of this model has not been on
cosmological systems (as contrasted to the matter-induced decoherence
of spacetime described in \cite{JJH89,padman} and the recent
minisuperspace work \cite{R&S}).  The background spacetime was taken
as Minkowski space and the temperature of the short-wavelength
graviton state was taken to be its present-day value.  Different
background spacetimes might also be studied once the tensor nature of
perturbative GR is restored.

\section*{Acknowledgments}

        The author would like to thank B.~K.~Berger, D.~A.~Craig,
J.~B.~Hartle, G.~T.~Horowitz, K.~V.~Kucha\v{r}, R.~H.~Price, and
J.~D.~Richman for advice and feedback, and also the Physics Department
of the University of California, Santa Barbara, where much of this
work was done.  This work was supported by NSF Grants PHY90-08502,
PHY95-07065, and PHY-9507719.


\appendix

\section{Glossary of notation}\label{app:gloss}

        For reference, we list here some of the notational conventions
and important symbols used throughout the body of the paper.

        $D$ is the number of \emph{space} dimensions, so the case of
physical interest is $D=3$.

        In the general discussion of Sec.~\ref{sec:infl}, the field
$\varphi$ is divided into a set of variables $\Phi$ which describe the
system and the remainder, $\phi$, which describe the environment.

        In the more concrete subsequent calculations, $\varphi(\x,t)$
is the field and $\varphi_\kay(t)$ its Fourier transform, defined by
\eqref{fourier}.\footnote{The arguments or subscripts are sometimes
omitted, in cases where the meaning of $\varphi$ (or $\Phi$ or $\phi$)
should be clear from context.}

        The space of wave vectors $\kay$ is divided into short- and
long-wavelength regions $\mc{L}$
and $\mc{S}$ illustrated in Fig.~\ref{fig:LS2}.  An arbitrarily-chosen
half $\mc{S}/2$ of the region $\mc{S}$ is also defined there.  When a
wave vector falls in the long-wavelength $\mc{L}$, it is conventional
to call it $\q$ rather than $\kay$.

        The Fourier modes $\varphi_\kay(t)$ are divided into
long-wavelength modes (LWMs) $\Phi_\q(t)$ and short-wavelength modes
(SWMs) $\phi_\kay(t)$ defined by \eqref{Phiphi}, making concrete the
formal system-environment split described
before.\footnote{Section~\protect\ref{ssec:imprac}, in which the roles of the
system and environment are reversed, is an exception to this.}

        The action $S[\varphi]$ is divided formally in
Sec.~\ref{sec:infl} into a system-only part $S_\Phi[\Phi]$ and an
interaction part $S_E[\phi,\Phi]$.  In the specific realization for
the scalar field theory, $S_E[\phi,\Phi]$ is divided into pieces
$S_0[\phi]$, $S_\phi[\phi,\Phi]$, $S_{\phi\phi}[\phi,\Phi]$, and
$S_\fff[\phi]$, as defined in \eqref{Lparts}.  Also of use are
$S_\off[\phi,\Phi]=S_0[\phi]+S_{\phi\phi}[\phi,\Phi]$ and
$S_3[\phi,\Phi]=S_\off[\phi,\Phi]+S_\phi[\phi]$.  Each of these
actions is defined by a corresponding Lagrangian $L[\phi,\Phi]$,
$L_\Phi[\Phi]$, etc.

        The influence phase $W[\Phi,\Phi']$ and influence functional
$e^{iW[\Phi,\Phi']}$ are defined by \eqref{infl}.  In addition,
corresponding phases $W_\off[\Phi,\Phi']$ and $W_3[\Phi,\Phi']$ are
defined by replacing $S_E[\phi,\Phi]$ with $S_\off[\phi,\Phi]$ and
$S_3[\phi,\Phi]$, respectively.

        The endpoints of the path $\varphi(\x,t)$ are indicated by
$\varphi_i(\x)=\varphi(\x,-T/2)$ and $\varphi_f(\x)=\varphi(\x,T/2)$.
Similarly, $\Phi_i$ and $\Phi_f$ are the endpoints of the path $\Phi$,
and $\phi_i$ and $\phi_f$ are the endpoints of the path $\phi$.

        A prime on a path or its endpoints is used to distinguish the
two arguments of the influence phase $W[\Phi,\Phi']$, or the
decoherence functional, as in $D[\varphi,\varphi']$.  The difference
between the two paths is written as in $\Delta\Phi=\Phi-\Phi'$ and
their average as in $\overline{\phi_i}=\frac{\phi_i+\phi'_i}{2}$.

        When an object is a functional of an entire path, its argument
is written in square brackets as in $S[\varphi]$.  When it depends
only on the value of the field at specific time, it is written, to
emphasize the distinction, as an ordinary function as in
$\rho(\varphi_i,\varphi_i')$, even though the argument is technically
still a function of the spatial \coord $\x$ or the mode label
$\kay$.\footnote{This notation is also justified in the case of, for
example, $\phi_i$, which may be thought of as a vector and not a
function.}  Similarly, the notation $\D\phi$ is reserved for path
integrals, with the functional integration over the endpoints written
as an ordinary integral as in $d\phi_i$.  The mismatched parentheses
in $\mc{K}_\off(\phi_f|\phi_i;\Phi]$ indicate that it is a
``function'' of the arguments before the semicolon and a functional of
the argument after it.

        The short-wavelength modes $\phi_\kay(t)$ are combined, by
\eqref{phivec}, into a vector $\phi$ in the space
$\R^{\mc{S}/2\oplus\mc{S}/2}$ described in Sec.~\ref{ssec:vec}.

        The following are all matrices which take vectors in
$\R^{\mc{S}/2\oplus\mc{S}/2}$ to other vectors in
$\R^{\mc{S}/2\oplus\mc{S}/2}$:\footnote{Note that in
Appendix~\ref{app:pert} there are equivalent matrices, defined
according to \eqref{ULR}, mapping $\R^{\mc{S}}$ to $\R^{\mc{S}}$.
This allows us to define, for instance, the components of
$\Brune{}_{1}$ as $\Brune{}_{1\M_1\M_2}$.} $m(t)$, defined in
\eqref{mdef}; $\varpi(t)$, defined in \eqref{varpidef};
$\Arune(t_f|t_i)$, $\Brune(t_f|t_i)$, and $\Crune(t_f|t_i)$, defined
in \eqref{solns}; $A[\Phi]$, $B[\Phi]$, and $C[\Phi]$, defined in
\eqref{ABC}; $\mc{A}_\pm$, $\mc{B}_\pm$, $\mc{C}_\pm$, defined in
\eqref{Apm}; and $\aleph_0$ and $\aleph_1[\Delta\Phi]$, defined in
\eqref{aleph}.

        $\mc{V}$ is a function having the character of its argument,
so $\mc{V}(k)$ is a number, and $\mc{V}(\Oo)$ is a matrix.

        In addition, $\mc{M}$ defined by \eqref{Mdef} and
$\wt{\mc{M}}$ defined by \eqref{Mtilde} are matrices which take
vectors in the product space $(\R^{\mc{S}/2\oplus\mc{S}/2})^3$ to
other vectors in $(\R^{\mc{S}/2\oplus\mc{S}/2})^3$.

        In most cases, a subscript of $0$ or $1$ indicates the zeroth-
or first-order contribution to the quantity in question, in an
expansion in powers of the coupling constant $\ell$.

        Finally, the field averages $\langle\,\rangle$ and
$\wt{\langle\,\rangle}$ are defined in \eqref{avedef} and \eqref{CG},
respectively.

\section{Perturbative expansions}\label{app:pert}

\subsection{$\Arune$, $\Brune$ and $\Crune$}\label{ssec:pertrune}

        To make practical use of the exact expressions \eqref{solns},
we need to expand them in powers of the coupling constant $\ell$.
Expanding to zeroth order is trivial, since $m_0=1$ and
$\varpi_0=\Oo^2$, where
\begin{equation}
\Oo=
\left(\begin{array}{cccc}
\{k_{1}\delta^\Ddim\!(\M_1-\M_2)\} & \{0\} \\
\{0\} & \{k_{1}\delta^\Ddim\!(\M_1-\M_2)\}
\end{array}\right)
\end{equation}
and of course
\begin{equation}
1=
\left(\begin{array}{cccc}
\{\delta^\Ddim\!(\M_1-\M_2)\} & \{0\} \\
\{0\} & \{\delta^\Ddim\!(\M_1-\M_2)\}
\end{array}\right).
\end{equation}
The expansion is
\begin{mathletters}\label{rune0}
\begin{eqnarray}
\Arune_0&=&\cos\Oo T=\Brune_0;
\\
\Crune_0&=&\frac{\sin\Oo T}{\Oo}\label{rune0C}.
\end{eqnarray}
\end{mathletters}

Proceeding to the first order terms, we can substitute
the first order expression
\begin{equation}
\prod_{k=n}^1\left[
        -m^{-1}(\tilde{t}_k)\varpi(t_k)
\right]
=1+\ell\sum_{k=1}^n 
(-\Oo^2)^{n-k}[\varpi_1(t_k)\Oo^{-2}-m_1(\tilde{t}_k)](-\Oo^2)^k
+\mc{O}(\ell^2)
\end{equation}
into \eqref{Bsoln} and find
\begin{eqnarray}
\Brune_1
=\sum_{n=1}^\infty
\frac{(-1)^n}{(2n-1)!}\int_{t_i}^{t_f}dt\,
\sum_{k=1}^n
\Biggl\{
&&      \Oo^{-1}[\Oo(t-t_i)]^{2n-2k+1}\tbinom{2n-1}{2k-2}
        \varpi_1(t)[\Oo(t_f-t)]^{2k-2}
\nonumber\\
&&
        -[\Oo(t-t_i)]^{2n-2k}\tbinom{2n-1}{2k-1}m_1(t)
        [\Oo(t_f-t)]^{2k-1}\Oo
\Biggr\}.
\end{eqnarray}

        To proceed further, we should streamline the notation for
components of matrices on $\R^{\mc{S}/2\oplus\mc{S}/2}$.  At the
moment, the components of $\Oo$ are written as
$\Omega^{\mathsf{UL}}_{0\M_1\M_2}=k_{1}\delta^\Ddim\!(\M_1-\M_2)$,
$\Omega^{\mathsf{UR}}_{0\M_1\M_2}=0$,
$\Omega^{\mathsf{LL}}_{0\M_1\M_2}=0$, and
$\Omega^{\mathsf{LR}}_{0\M_1\M_2}=k_{1}\delta^\Ddim\!(\M_1-\M_2)$.  This is
greatly simplified if we observe that
$\abs{-\M}=k$, and define
\begin{equation}\label{ULR}
M=\left(\begin{array}{cc}
\{M^{\mathsf{UL}}_{\M_1\M_2}\} & \{M^{\mathsf{UR}}_{\M_1\M_2}\}
\\
\{M^{\mathsf{LL}}_{\M_1\M_2}\} & \{M^{\mathsf{LR}}_{\M_1\M_2}\}
\end{array}\right)
=\left(\begin{array}{cc}
\{M_{\M_1\M_2}\} & \{M_{\M_1,-\M_2}\}
\\
\{M_{-\M_1,\M_2}\} & \{M_{-\M_1,-\M_2}\}
\end{array}\right)
=\{M_{\M_1\M_2}\},
\end{equation}
where the indices in the first two expressions range over
$\mc{S}/2$ and those in the third range over $\mc{S}$.  Then
$\Omega_{0\M_1\M_2}=k_{1}\delta^\Ddim\!(\M_1-\M_2)$, and
\begin{eqnarray}
\Brune{}_{1\M_1\M_2}&=&\sum_{n=1}^\infty
\frac{(-1)^n}{(2n-1)!}\int_{t_i}^{t_f}dt\,
\left\{
        \frac{\varpi_1(t)_{\M_1\M_2}}{k_{1}}
        \sum_{k=1}^n\tbinom{2n-1}{2k-2}
        [\overbrace{k_{1}(t-t_i)}^{\theta_{A1}}]^{2n-2k+1}
        [\overbrace{k_{2}(t_f-t)}^{\theta_{B2}}]^{2k-2}
\right.
\nonumber\\
&&\phantom{
  \sum_{n=1}^\infty
  \frac{(-1)^n}{(2n-1)!}\int_{t_i}^{t_f}dt\,
  \Biggl\{
}
\left.
        -k_{2}m_1(t)_{\M_1\M_2}
        \sum_{k=1}^n\tbinom{2n-1}{2k-1}
        [k_{1}(t-t_i)]^{2n-2k}[k_{2}(t_f-t)]^{2k-1}
\right\}
\nonumber\\
&=&\int_{t_i}^{t_f}dt\,
\left[
        m_1(t)_{\M_1\M_2}k_{2}\cos\theta_{A1}\sin\theta_{B2}
        -\varpi_1(t)_{\M_1\M_2}k_{1}^{-1}
        \sin\theta_{A1}\cos\theta_{B2}
\right]\label{B1}
\end{eqnarray}

        Then we can use the first order term in \eqref{Csoln},
\begin{equation}
\Crune_1=-\frac{d\Brune_1}{dt_f}\Oo^{-2}
-\Oo\sin\Oo T\Oo^{-2}\varpi_1(t_f)\Oo^{-2},
\end{equation}
to calculate
\begin{equation}
\Crune_{1\M_1\M_2}=-\int_{t_i}^{t_f}dt\,
\left[
        m_1(t)_{\M_1\M_2}\cos\theta_{A1}\cos\theta_{B2}
\label{C1}
        +\varpi_1(t)_{\M_1\M_2}
        \frac{\sin\theta_{A1}\sin\theta_{B2}}{k_{1}k_{2}}
\right];
\end{equation}
likewise, the first order term in \eqref{Asoln},
\begin{equation}
\Arune_1=-\frac{d\Crune_1}{dt_i}+m_1(t_i)\cos\Oo T,
\end{equation}
gives
\begin{equation}
\Arune_{1\M_1\M_2}=\int_{t_i}^{t_f}dt\,
\left[
        m_1(t)_{\M_1\M_2}k_{1}\sin\theta_{A1}\cos\theta_{B2}
\label{A1}
        -\varpi_1(t)_{\M_1\M_2}k_{2}^{-1}
        \cos\theta_{A1}\sin\theta_{B2}
\right].
\end{equation}
It is straightforward to check that \eqref{B1}, \eqref{C1}, and
\eqref{A1} satisfy \eqref{dtb} and \eqref{dta}

\subsection{$A$, $B$ and $C$}\label{ssec:pert}

We can use the expansions for $\Arune$, $\Brune$ and $\Crune$ calculated
in Section~\ref{ssec:pertrune} to find expansions for
\begin{eqnarray}
A[\Phi]&=&\Arune\left(\frac{T}{2}\right.\left|-\frac{T}{2}\right)
\Crune^{-1}\left(\frac{T}{2}\right.\left|-\frac{T}{2}\right)
-\dot{m}\left(-\frac{T}{2}\right),\tag{\ref{APdef}}
\\
B[\Phi]&=&\Crune^{-1}\left(\frac{T}{2}\right.\left|-\frac{T}{2}\right)
\Brune\left(\frac{T}{2}\right.\left|-\frac{T}{2}\right)
+\dot{m}\left(\frac{T}{2}\right),\tag{\ref{BPdef}}
\\
C[\Phi]&=&\Crune^{-1}\left(\frac{T}{2}\right.\left|-\frac{T}{2}\right).
\tag{\ref{CPdef}}
\end{eqnarray}

        {From} \eqref{rune0}, the zero order terms are
\begin{mathletters}
\begin{eqnarray}
A_0&=&\frac{\Oo}{\tan\Oo T}=B_0
\\
C_0&=&\frac{\Oo}{\sin\Oo T};
\end{eqnarray}
\end{mathletters}
Proceeding to the first order terms, we have
\begin{equation}
A_1[\Phi]
=\Arune_1\frac{\Oo}{\sin\Oo T}
-\cos\Oo T\frac{\Oo}{\sin\Oo T}\Crune_1\frac{\Oo}{\sin\Oo T}
-\dot{m}_1\left(-\frac{T}{2}\right).
\end{equation}
Using \eqref{A1} and \eqref{C1} 
and rewriting the $\dot{m}$ boundary term using
\begin{equation}
-\dot{m}_1\left(-\frac{T}{2}\right)_{\M_1\M_2}\sin k_{1}T\sin k_{2}T
=\dot{m}(t)_{\M_1\M_2}\sin\theta_{B1}\sin\theta_{B2}
\bigr\rvert_{-T/2}^{T/2},
\end{equation}
we can calculate, via integration by parts,
\begin{eqnarray}
A_{1\M_1\M_2}&=&m_1\left(-\frac{T}{2}\right)_{\M_1\M_2}
\left(\frac{k_{1}}{\tan k_{1}T}+\frac{k_{2}}{\tan k_{2}T}\right)
\nonumber\\
&&-\frac{k_{1}k_{2}}{\sin k_{1}T\sin k_{2}T}
\int_{-T/2}^{T/2}dt\,
\biggl[
        m_1(t)_{\M_1\M_2}\cos\theta_{B1}\cos\theta_{B2}
        +n_1(t)_{\M_1\M_2}\sin\theta_{B1}\sin\theta_{B2}
\biggr],\label{An}
\end{eqnarray}
where
\begin{equation}\label{ndef}
n_1(t)_{\M_1\M_2}=\frac{\varpi_1(t)_{\M_1\M_2}-\ddot{m}_1(t)_{\M_1\M_2}
-m_1(t)_{\M_1\M_2}(k_{1}^2+k_{2}^2)}{k_{1}k_{2}}.
\end{equation}
Analogously, we find
\begin{eqnarray}
B_{1\M_1\M_2}&=&m_1\left(\frac{T}{2}\right)_{\M_1\M_2}
\left(\frac{k_{1}}{\tan k_{1}T}+\frac{k_{2}}{\tan k_{2}T}\right)
\nonumber\\
&&-\frac{k_{1}k_{2}}{\sin k_{1}T\sin k_{2}T}
\int_{-T/2}^{T/2}dt\,
\biggl[
        m_1(t)_{\M_1\M_2}\cos\theta_{A1}\cos\theta_{A2}
        +n_1(t)_{\M_1\M_2}\sin\theta_{A1}\sin\theta_{A2}
\biggr].\label{Bn}
\end{eqnarray}
The first order term $C_1=-\frac{\Oo}{\sin\Oo T}
\Crune_1\frac{\Oo}{\sin\Oo T}$ can be cast into a similar form
after integration by parts to give
\begin{eqnarray}
C_{1\M_1\M_2}&=&m_1\left(\frac{T}{2}\right)_{\M_1\M_2}\frac{k_{2}}{\sin k_{2}T}
+m_1\left(-\frac{T}{2}\right)_{\M_1\M_2}\frac{k_{1}}{\sin k_{1}T}
\nonumber\\
&&-\frac{k_{1}k_{2}}{\sin k_{1}T\sin k_{2}T}
\int_{-T/2}^{T/2}dt\,
\biggl[
        m_1(t)_{\M_1\M_2}\cos\theta_{A1}\cos\theta_{B2}
        -n_1(t)_{\M_1\M_2}\sin\theta_{A1}\sin\theta_{B2}
\biggr].\label{Cn}
\end{eqnarray}
Application of trigonometric identities to expressions such as
$\cos\theta_{A1}\cos\theta_{B2}=\cos k_{1}(t+T/2)\cos k_{2}(T/2-t)$
allows us to rewrite \eqref{An}, \eqref{Bn} and \eqref{Cn} in terms of
$k_{\pm}=k_{1}\pm k_{2}$ as
\begin{mathletters}
\begin{eqnarray}
A_{1\M_1\M_2}&=&m_1\left(-\frac{T}{2}\right)_{\M_1\M_2}
\left(\frac{k_{1}}{\tan k_{1}T}+\frac{k_{2}}{\tan k_{2}T}\right)
\nonumber\\
&&-\frac{k_{1}k_{2}}{2\sin k_{1}T\sin k_{2}T}
\int_{-T/2}^{T/2}dt\,
\left\{
        [m_1(t)+n_1(t)]_{\M_1\M_2}\left(
                \cos k_{-}t\cos\frac{k_{-}T}{2}
                +\sin k_{-}t\sin\frac{k_{-}T}{2}
        \right)
\right.
\nonumber\\
&&\phantom{-\frac{k_{1}k_{2}}{2\sin k_{1}T\sin k_{2}T}
\int_{-T/2}^{T/2}dt\,\biggl\{
}\left.
        +[m_1(t)-n_1(t)]_{\M_1\M_2}\left(
                \cos k_{+}t\cos\frac{k_{+}T}{2}
                +\sin k_{+}t\sin\frac{k_{+}T}{2}
        \right)
\right\}
\\
B_{1\M_1\M_2}&=&m_1\left(\frac{T}{2}\right)_{\M_1\M_2}
\left(\frac{k_{1}}{\tan k_{1}T}+\frac{k_{2}}{\tan k_{2}T}\right)
\nonumber\\
&&-\frac{k_{1}k_{2}}{2\sin k_{1}T\sin k_{2}T}
\int_{-T/2}^{T/2}dt\,
\left\{
        [m_1(t)+n_1(t)]_{\M_1\M_2}\left(
                \cos k_{-}t\cos\frac{k_{-}T}{2}
                -\sin k_{-}t\sin\frac{k_{-}T}{2}
        \right)
\right.
\nonumber\\
&&\phantom{-\frac{k_{1}k_{2}}{2\sin k_{1}T\sin k_{2}T}
\int_{-T/2}^{T/2}dt\,\biggl\{
}\left.
        +[m_1(t)-n_1(t)]_{\M_1\M_2}\left(
                \cos k_{+}t\cos\frac{k_{+}T}{2}
                -\sin k_{+}t\sin\frac{k_{+}T}{2}
        \right)
\right\}
\\
C_{1\M_1\M_2}&=&m_1\left(\frac{T}{2}\right)_{\M_1\M_2}\frac{k_{2}}{\sin k_{2}T}
+m_1\left(-\frac{T}{2}\right)_{\M_1\M_2}\frac{k_{1}}{\sin k_{1}T}
\nonumber\\
&&-\frac{k_{1}k_{2}}{2\sin k_{1}T\sin k_{2}T}
\int_{-T/2}^{T/2}dt\,
\left\{
        [m_1(t)+n_1(t)]_{\M_1\M_2}\left(
                \cos k_{-}t\cos\frac{k_{+}T}{2}
                -\sin k_{-}t\sin\frac{k_{+}T}{2}
        \right)
\right.
\nonumber\\
&&\phantom{-\frac{k_{1}k_{2}}{2\sin k_{1}T\sin k_{2}T}
\int_{-T/2}^{T/2}dt\,\biggl\{
}\left.
        +[m_1(t)-n_1(t)]_{\M_1\M_2}\left(
                \cos k_{+}t\cos\frac{k_{-}T}{2}
                -\sin k_{+}t\sin\frac{k_{-}T}{2}
        \right)
\right\}
\end{eqnarray}
\end{mathletters}
        These seem to be taking on a nice form in terms of more
basically defined modes
\begin{mathletters}
\begin{eqnarray}
\chi_{\pm}&=&\int_{-T/2}^{T/2}dt\,[m_1(t)\mp n_1(t)]_{\M_1\M_2}\cos k_{\pm}t
+\text{boundary terms},
\\
\sigma_{\pm}&=&\int_{-T/2}^{T/2}dt\,[m_1(t)\mp n_1(t)]_{\M_1\M_2}\sin k_{\pm}t
+\text{boundary terms},
\end{eqnarray}
\end{mathletters}
namely\footnote{Note that the elements in these matrices are numbers,
rather than matrices, {i.e.}, the expression holds for a
particular value of $\kay_1$ and $\kay_2$ so that there is no integral
over $\kay_2$ included in the matrix multiplication.}
\begin{equation}\label{chisigxf}
\left(\begin{array}{c}
B_{1\M_1\M_2} \\ A_{1\M_1\M_2} \\ C_{1\M_1\M_2} \\ \trans{C_{1}}_{\M_1\M_2}
\end{array}\right)
=
-\frac{k_{1}k_{2}}{2\sin k_{1}T\sin k_{2}T}
\left(\begin{array}{cccc}
\cos\frac{k_{-}T}{2} & -\sin\frac{k_{-}T}{2}
& \cos\frac{k_{+}T}{2} & -\sin\frac{k_{+}T}{2} \\
\cos\frac{k_{-}T}{2} & \sin\frac{k_{-}T}{2}
& \cos\frac{k_{+}T}{2} & \sin\frac{k_{+}T}{2} \\
\cos\frac{k_{+}T}{2} & -\sin\frac{k_{+}T}{2}
& \cos\frac{k_{-}T}{2} & -\sin\frac{k_{-}T}{2} \\
\cos\frac{k_{+}T}{2} & \sin\frac{k_{+}T}{2}
& \cos\frac{k_{-}T}{2} & \sin\frac{k_{-}T}{2}
\end{array}\right)
\!\!
\left(\begin{array}{c}
\chi_{-} \\ \sigma_{-} \\ \chi_{+} \\ \sigma_{+}
\end{array}\right),
\end{equation}
where of course $\trans{C_{1}}_{\M_1\M_2}=C_{1\M_2\M_1}$.  We can
use this to define $\chi_\pm$ and $\sigma_\pm$, and inverting the
matrix 
determines the boundary terms, giving
\begin{mathletters}
\begin{eqnarray}
\chi_{\pm}&=&\int_{-T/2}^{T/2}dt\,[m_1(t)\mp n_1(t)]_{\M_1\M_2}\cos k_{\pm}t
\mp\left.m_1(t)_{\M_1\M_2}\frac{k_{\pm}}{k_{1}k_{2}}
\sin 2k_{\pm}t\right\rvert_{-T/2}^{T/2},\label{chipm}
\\
\sigma_{\pm}&=&\int_{-T/2}^{T/2}dt\,[m_1(t)\mp n_1(t)]_{\M_1\M_2}\sin k_{\pm}t
\pm\left.m_1(t)_{\M_1\M_2}\frac{k_{\pm}}{k_{1}k_{2}}
\cos 2k_{\pm}t\right\rvert_{-T/2}^{T/2}.\label{sigpm}
\end{eqnarray}
\end{mathletters}

\section{Evaluation of the trace in
Sec.~\protect\ref{ssec:eval}}\label{app:aleph}

The purpose of this appendix is to insert the values for $A_1$, $B_1$
and $C_1$ from Sec.~\ref{ssec:pert} of Appendix~\ref{app:pert} into
the expression
$E^2[\Delta\Phi]$ appearing in
the limit \eqref{decTR} on the influence phase, where
\begin{eqnarray}
&&\aleph_0=\frac{\Oo^2\mc{V}(\Oo)}{\sin^2\Oo T}
\left(\begin{array}{cc}
1 & -\cos\Oo T \\ -\cos\Oo T & 1
\end{array}\right)
\tag{\ref{alepho}}
\\
\tag{\ref{alephi}}
&&\aleph_1[\Delta\Phi]=
\left(\begin{array}{cc}
B_1[\Delta\Phi] & -C_1[\Delta\Phi] \\
-\trans{C_1[\Delta\Phi]} & A_1[\Delta\Phi]
\end{array}\right)
\end{eqnarray}

Inverting $\aleph_0$ gives
\begin{equation}
\aleph_0^{-1}\aleph_1
=\frac{1}{\Oo^2\mc{V}(\Oo)}
\left(\begin{array}{cccc}
B_1 - \cos\Oo T \trans{C_1}  &  -C_1 + \cos\Oo T A_1 \\
\cos\Oo T B_1 - \trans{C_1}  &  -\cos\Oo T C_1 + A_1
\end{array}\right)
\end{equation}
so\footnote{For the conversion of the range of the indices of these
real matrices from $\mc{S}/2\oplus\mc{S}/2$ to $\mc{S}$, see
\eqref{ULR}.}
\begin{eqnarray}
E^2[\Delta\Phi]
&=&\Tr\left(\ell\aleph_0^{-1}\aleph_1\right)^2
=\int\limits_{\mc{S}} 
\frac{\ell^2 d^\Ddim\!k_1\, d^\Ddim\!k_2}
{k_{1}^2\mc{V}(k_{1})k_{2}^2\mc{V}(k_{2})}
\nonumber\\
&&\times
\trans{
 \left(\begin{array}{cccc}
 B_{1\M_2\M_1} \\ A_{1\M_2\M_1} \\ \trans{C_{1}}_{\M_2\M_1} \\ C_{1\M_2\M_1}
 \end{array}\right)
}
\left(\begin{array}{cccc}
 1 & \cos k_{1}T\cos k_{2}T & -\cos k_{2}T & -\cos k_{1}T \\
 \cos k_{1}T\cos k_{2}T & 1 & -\cos k_{1}T & -\cos k_{2}T \\
 -\cos k_{2}T & -\cos k_{1}T & 1 & \cos k_{1}T\cos k_{2}T \\
 -\cos k_{1}T & -\cos k_{2}T & \cos k_{1}T\cos k_{2}T & 1
\end{array}\right)
\left(\begin{array}{cccc}
 B_{1\M_1\M_2} \\ A_{1\M_1\M_2} \\ C_{1\M_1\M_2} \\ \trans{C_{1}}_{\M_1\M_2}
\end{array}\right).
\end{eqnarray}
Using the result \eqref{chisigxf} from Sec.~\ref{ssec:pert} of
Appendix~\ref{app:pert} and performing the matrix multiplication gives
\begin{equation}
E^2[\Delta\Phi]
=\int\limits_{\mc{S}} 
\frac{\ell^2 d^\Ddim\!k_1\, d^\Ddim\!k_2}
{4\mc{V}(k_{1})\mc{V}(k_{2})}
\left(\chi_{-}^2+\sigma_{-}^2+\chi_{+}^2+\sigma_{+}^2\right).
\end{equation}

        Now it's time to take the result in terms of the real matrices
$m_1[\Phi]$ and $\varpi_1[\Phi]$ on $\R^{\mc{S}/2\oplus\mc{S}/2}$
defined by \eqref{mdef} and
\eqref{varpidef}, and reconstruct from them useful expressions in
terms of $\{\Phi_\N\}$ and the complex matrices on $\C^{\mc{S}}$ 
with elements
\begin{mathletters}
\begin{eqnarray}
(m_{1\text{ex}})_{\M_1\M_2}&=&\Phi_{\M_1-\M_2}\label{m1ex}
\\
(\varpi_{1\text{ex}})_{\M_1\M_2}&=&
k^2_{12}(m_{1\text{ex}})_{\M_1\M_2}
+(\ddot{m}_{1\text{ex}})_{\M_1\M_2}
\end{eqnarray}
defined by \eqref{mex}, as well as [{cf.}\
\eqref{ndef}]
\begin{eqnarray}
(n_{1\text{ex}})_{\M_1\M_2}
&=&
\frac{(\varpi_{1\text{ex}})_{\M_1\M_2}-(\ddot{m}_{1\text{ex}})_{\M_1\M_2}
-(m_{1\text{ex}})_{\M_1\M_2}(k_{1}^2+k_{2}^2)}{k_{1}k_{2}}
=\frac{(m_{1\text{ex}})_{\M_1\M_2}(k^2_{12}-k_{1}^2-k_{2}^2)}
{k_{1}k_{2}}
\nonumber\\
&=&-(m_{1\text{ex}})_{\M_1\M_2}\frac{\kay_{1}\cdot\kay_{2}}
{k_{1}k_{2}}
=-(m_{1\text{ex}})_{\M_1\M_2}\cos\theta_{12}.\label{n1ex}
\end{eqnarray}
\end{mathletters}
These are related to the real matrices $m$ and $\varpi$ on $\R^\mc{S}$
(or $\R^{\mc{S}/2\oplus\mc{S}/2}$) as follows: for
$\M_1,\M_2\in\mc{S}/2$,
\begin{mathletters}
\begin{eqnarray}
M^{\mathsf{UL}}_{\M_1\M_2}&=&M_{\M_1\M_2}
=(\Mex)^{\mathrm{R}}_{\M_1\M_2}+(\Mex)^{\mathrm{R}}_{1,-\M_2}
\\
M^{\mathsf{LR}}_{\M_1\M_2}&=&M_{-\M_1,-\M_2}
=(\Mex)^{\mathrm{R}}_{\M_1\M_2}-(\Mex)^{\mathrm{R}}_{1,-\M_2}
\\
M^{\mathsf{LL}}_{\M_1\M_2}&=&M_{-\M_1\M_2}
=(\Mex)^{\mathrm{I}}_{\M_1\M_2}+(\Mex)^{\mathrm{I}}_{1,-\M_2}
\\
M^{\mathsf{UR}}_{\M_1\M_2}&=&M_{1,-\M_2}
=-(\Mex)^{\mathrm{I}}_{\M_1\M_2}+(\Mex)^{\mathrm{I}}_{1,-\M_2}.
\end{eqnarray}
\end{mathletters}
Since
\begin{eqnarray}
\int\limits_{\mc{S}} d^\Ddim\!k_1\, d^\Ddim\!k_2\,
\left(M_{\M_1\M_2}\right)^2
&=&2\int\limits_{\mc{S}/2} d^\Ddim\!k_1\, d^\Ddim\!k_2\,
\left\{
        \left[(\Mex)^{\mathrm{R}}_{\M_1\M_2}\right]^2
        +\left[(\Mex)^{\mathrm{R}}_{1,-\M_2}\right]^2
        +\left[(\Mex)^{\mathrm{I}}_{\M_1\M_2}\right]^2
        +\left[(\Mex)^{\mathrm{I}}_{1,-\M_2}\right]^2
\right\}
\nonumber\\
&=&\int\limits_{\mc{S}} d^\Ddim\!k_1\, d^\Ddim\!k_2\,
\abs{\left(\Mex\right)_{\M_1\M_2}}^2,
\end{eqnarray}
we can use \eqref{m1ex} and \eqref{n1ex}, along with
$\abs{\sum_i\alpha_i\beta_i^{\mathrm{R}}}^2
+\abs{\sum_i\alpha_i\beta_i^{\mathrm{I}}}^2$
$=\frac{1}{2}\left(\abs{\sum_i\alpha_i\beta_i}^2\right.$
$\left.+\abs{\sum_i\alpha_i\beta_i^*}^2\right)$, to write
\begin{eqnarray}
E^2[\Delta\Phi]=\int\limits_{k_1,k_2>k_c}
\frac{d^\Ddim\!k_1\,d^\Ddim\!k_2\,\Theta(k_c-q)}
{4\mc{V}(k_1)\mc{V}(k_2)}&&
\left\{
\vphantom{\abs{\int_{-T/2}^{T/2}}^2}
\right.
        \abs{\sin^2\frac{\theta_{12}}{2}
        \int_{-T/2}^{T/2}dt\,\ell\Delta\Phi_\q(t)e^{ik_{-}t}
        -i\frac{k_{-}}{k_1k_2}
        \left[e^{i2k_{-}t}\ell\Delta\Phi_\q(t)
        \right]_{-T/2}^{T/2}}^2
\nonumber\\
        &&+\abs{\sin^2\frac{\theta_{12}}{2}
        \int_{-T/2}^{T/2}dt\,\ell\Delta\Phi_\q(t)e^{-ik_{-}t}
        +i\frac{k_{-}}{k_1k_2}
        \left[e^{-i2k_{-}t}\ell\Delta\Phi_\q(t)
        \right]_{-T/2}^{T/2}}^2
\nonumber\\
        &&+\abs{\cos^2\frac{\theta_{12}}{2}
        \int_{-T/2}^{T/2}dt\,\ell\Delta\Phi_\q(t)e^{ik_{+}t}
        +i\frac{k_{+}}{k_1k_2}
        \left[e^{i2k_{+}t}\ell\Delta\Phi_\q(t)
        \right]_{-T/2}^{T/2}}^2
\tag{\ref{TRcont}}\\ \nonumber
        &&+\abs{\cos^2\frac{\theta_{12}}{2}
        \int_{-T/2}^{T/2}dt\,\ell\Delta\Phi_\q(t)e^{-ik_{+}t}
        -i\frac{k_{+}}{k_1k_2}
        \left[e^{-i2k_{+}t}\ell\Delta\Phi_\q(t)
        \right]_{-T/2}^{T/2}}^2
\left.
\vphantom{\abs{\int_{-T/2}^{T/2}}^2}
\right\}.
\end{eqnarray}

\section{The effects of the linear and cubic terms}
\label{app:full}

        Here we 
add the terms $S_\phi$ and $S_\fff$ back into the action, and
determine what effect, if any, this has on the influence phase
\eqref{decTR}.

\subsection{The linear terms}\label{ssec:line}

         The effect of
the linear term $S_\phi$ can, as usual, be elucidated by completing
the square, as shown in this section.

        We define the ``all-but-cubic'' Lagrangian
\begin{equation}
L_3[\phi,\Phi]=L_\off[\phi,\Phi]+\ell L_\phi[\phi,\Phi]
\end{equation}
by adding to the quadratic action considered in
Sec.~\ref{sec:quad} the linear terms
\begin{equation}
\ell L_\phi[\phi,\Phi]=\int\limits_{\mc{S}}d^\Ddim\!k\,
\left(-\phi_\M^*\wt{x}_\M + \dot{\phi}_\M^*\wt{y}_\M\right)
\end{equation}
where\footnote{Recall that $\Phi_{\M-\N}=0$ when $\M-\N\notin\mc{L}$}
[{cf.}\ \eqref{Lphi}]
\begin{mathletters}\label{xydef}
\begin{eqnarray}
\wt{x}_\M&=&-\frac{\ell}{2}\int\limits_{\mc{S}}d^\Ddim\!q\left(
        Q^2\Phi_{\M-\N}\Phi_\N
        -\dot{\Phi}_{\M-\N}\dot{\Phi}_\N
\right)
\\
\wt{y}_\M&=&-\frac{\ell}{2}\int\limits_{\mc{S}}d^\Ddim\!q\left(
        \dot{\Phi}_{\M-\N}\Phi_\N
        -\Phi_{\M-\N}\dot{\Phi}_\N
\right),
\end{eqnarray}
\end{mathletters}
with $Q^2$ bearing the same relation to $-\M$ and $\N$ that
$k^2_{12}$ [{cf.}\ \eqref{k2M}] bore to $\M_1$ and $\M_2$:
\begin{equation}
Q^2
=-(-\M)\cdot(\M-\N)
-(\M-\N)\cdot\N-\N\cdot(-\M)
=\M^2+\N^2-\M\cdot\N.
\end{equation}
The reality condition $\Phi_{-\N}=\Phi_\N^*$ forces
$\wt{x}_{-\M}=\wt{x}_\M^*$ and
$\wt{y}_{-\M}=\wt{y}_\M^*$, so we can use the identity
\begin{equation}
{v^{\text{ex}}}^\dagger w^{\text{ex}}
=\int\limits_{\mc{S}}d^\Ddim\!k\,v_\M^* w_\M
=\int\limits_{\mc{S}}d^\Ddim\!k\,\left(
        v_\M^{\mathrm{R}} w_\M^{\mathrm{R}}
        +v_\M^{\mathrm{I}} w_\M^{\mathrm{I}}
\right)
=2\int\limits_{\mc{S}/2}d^\Ddim\!k\,\left(
        v_\M^{\mathrm{R}} w_\M^{\mathrm{R}}
        +v_\M^{\mathrm{I}} w_\M^{\mathrm{I}}
\right)
=\trans{v}w,
\end{equation}
where $v$ and $w$ are vectors in $\R^{\mc{S}/2\oplus\mc{S}/2}$ defined
as in \eqref{phidef}, to write
$\ell L_\phi=-\trans{\phi}\wt{x}
+\trans{\dot{\phi}}\wt{y}$;
by integrating by parts, we can also write this, for an arbitrary
$z(t)$ as
\begin{equation}\label{Lphiz}
\ell L_\phi=-\trans{\phi}x+\trans{\dot{\phi}}y
+\frac{d}{dt}\left(\trans{\phi}z\right),
\end{equation}
where
\begin{mathletters}
\begin{eqnarray}
x&=&\wt{x}+\dot{z},\label{xdefz}
\\
y&=&\wt{y}-z,
\end{eqnarray}
\end{mathletters}
to give
\begin{equation}
L_3[\phi,\Phi]=\frac{1}{2}\left[\vphantom{\frac{d}{dt}}
        \trans{\dot{\phi}}m\dot{\phi}+2\trans{\dot{\phi}}y
        -\trans{\phi}\varpi\phi-2\trans{\phi}x
        +\frac{d}{dt}
        \left(\trans{\phi}\dot{m}\phi+2\trans{\phi}z\right)
\right].
\end{equation}
If we can choose $z$ such that
\begin{equation}\label{xyDE}
y=m\frac{d}{dt}(\varpi^{-1}x),
\end{equation}
$L_3$ is related to $L_\off$ by
\begin{equation}
L_{\off}[\phi+\varpi^{-1}x]
=L_3[\phi,\Phi]
+\frac{\trans{y}m^{-1}y}{2}-\frac{\trans{x}\varpi^{-1}x}{2}
+\frac{d}{dt}\left[\trans{\phi}(\dot{m}\varpi^{-1}x-z)\right]
+\frac{d}{dt}\frac{\trans{x}\varpi^{-1}\dot{m}\varpi^{-1}x}{2}.
\end{equation}
The condition \eqref{xyDE} is equivalent to the second order
inhomogeneous ODE
\begin{equation}\label{zDE}
\frac{d}{dt}(\varpi^{-1}\dot{z})+m^{-1}z
=m^{-1}\wt{y}-\frac{d}{dt}(\varpi^{-1}\wt{x}).
\end{equation}
A particular $\Phi(t)$ generates $\wt{x}(t)$ and
$\wt{y}(t)$ via \eqref{xydef}, and for that source term, we can
solve \eqref{zDE}, with the freedom to fix two boundary conditions
which are functions of $z_i$, $z_f$, $(\dot{z})_i$ and $(\dot{z})_f$.

        We can use this expression for $L_3$ to express $\mc{K}_3$,
the propagator for $S_3$, in terms of $\mc{K}_\off$ as
\begin{equation}
\mc{K}_3(\phi_f|\phi_i;\Phi]
\label{K3X}
=\mc{K}_\off(\phi_f|\phi_i;\Phi]
\exp\left\{
        i\trans{\left(\begin{array}{cccc}\phi_f\\ \phi_i\end{array}\right)}
        \mc{X}[\Phi]
\right\}
e^{i\psi[\Phi]},
\end{equation}
where
\begin{equation}\label{psidef}
\psi[\Phi]=\frac{1}{2}\int_{-T/2}^{T/2}dt\,(\trans{x}\varpi^{-1}x
-\trans{y}m^{-1}y)
+\frac{1}{2}
\trans{
 \left(\begin{array}{cccc}
        (\varpi^{-1}x)_f\\ (\varpi^{-1}x)_i
  \end{array}\right)
}
\left(\begin{array}{cccc}
        B[\Phi]-\dot{m}_f & -C[\Phi] \\
        -\trans{C[\Phi]}  & A[\Phi]+\dot{m}_i
\end{array}\right)
\left(\begin{array}{cccc}
(\varpi^{-1}x)_f\\ (\varpi^{-1}x)_i
\end{array}\right)
\end{equation}
is a real phase, and
\begin{equation}\label{Xdef}
\mc{X}[\Phi]=
\left(\begin{array}{cccc}
        z_f-(\dot{m}\varpi^{-1}x)_f\\
        -z_i+(\dot{m}\varpi^{-1}x)_i
\end{array}\right)
+\left(\begin{array}{cccc}
        B[\Phi] & -C[\Phi] \\
        -\trans{C[\Phi]} & A[\Phi]
\end{array}\right)
\left(\begin{array}{cccc}
        (\varpi^{-1}x)_f\\ (\varpi^{-1}x)_i
\end{array}\right).
\end{equation}
By substituting for $x$ using \eqref{xdefz}, we see that
$\mc{X}[\Phi]=0$ is just a pair of linear first order boundary
conditions on $z(t)$, and so we can choose the solution to \eqref{zDE}
to obey them, leaving
\begin{equation}\label{K3}
\mc{K}_3(\phi_f|\phi_i;\Phi]
=\mc{K}_\off(\phi_f|\phi_i;\Phi]
e^{i\psi[\Phi]}.
\end{equation}
Proceeding along the same lines as \eqref{WoffK}, we find that
\begin{equation}
e^{iW_3[\Phi,\Phi']}=e^{iW_\off[\Phi,\Phi']}
e^{i(\psi[\Phi]-\psi[\Phi'])}.
\end{equation}
Since $\psi[\Phi]$ is real, and it is the imaginary part
of $W$ which imposes decoherence, 
\begin{equation}\tag{\ref{threesame}}
\abs{e^{iW_3[\Phi,\Phi']}}=\abs{e^{iW_\off[\Phi,\Phi']}}
\end{equation}
and adding in the linear terms does not change the result
\eqref{decTR}.

\subsection{The cubic terms}\label{ssec:cube}

        Now we are ready to consider the full environmental action
\begin{equation}
S_E[\phi,\Phi]=S_3[\phi,\Phi]+S_\fff[\phi]
\end{equation}
including the cubic terms from
\begin{equation}
L_\fff[\phi]=
-\frac{1}{2}\int\limits_{\mc{S}}
d^\Ddim\!k_1\, d^\Ddim\!k_2\, d^\Ddim\!k_3\,
\delta^\Ddim\!(\M_1+\M_2+\M_3)
\left(
        \phi_{1}\dot{\phi}_{2}\dot{\phi}_{3}
        -\frac{\kay_{1}^2+\kay_{2}^2
        +\kay_{3}^2}{6}\phi_{1}\phi_{2}\phi_{3}
\right).
\end{equation}
Here we need to resort to using a generating functional
\begin{mathletters}
\begin{equation}\label{Zdef}
Z[J,J',\Phi,\Phi']=\int\D\phi\,\D\phi' 
\rho_\phi(\phi_i,\phi'_i)\delta(\phi'_f-\phi_f)
\exp\left\{i\left(
        S_3[\phi,\Phi]-S_3[\phi',\Phi']
        +\int_{-T/2}^{T/2}dt\, [\trans{\phi}J-\trans{\phi'}J']
\right)\right\}
\end{equation}
and expressing\footnote{The choice of sign of $J'$ may seem unusual,
but it allows us to write the argument of the exponential in
\eqref{Zdef} as $S_Z[\phi,\Phi,J]-S_Z[\phi',\Phi',J']$ rather
than $S_Z[\phi,\Phi,J]-S_Z[\phi',\Phi',-J']$.}
\begin{equation}\label{degen}
e^{iW[\Phi,\Phi']}=\exp\left(
        i\ell S_\fff\left[\frac{1}{i}\frac{\D}{\D J}\right]
        -i\ell S_\fff\left[-\frac{1}{i}\frac{\D}{\D J'}\right]
\right)
Z[J,J',\Phi,\Phi'].
\end{equation}
\end{mathletters}

\subsubsection{The wrong way to complete the square}

        The ``propagators'' involved in constructing the
generating functional 
\begin{equation}
Z[J,J',\Phi,\Phi']=\int d\phi_i d\phi'_i d\phi_f 
\rho_\phi(\phi_i,\phi'_i)
\mc{K}_Z(\phi_f|\phi_i;\Phi,J]
\mc{K}^*_Z(\phi_f|\phi'_i;\Phi',J']
\end{equation}
are of the form
\begin{equation}
\mc{K}_Z(\phi_f|\phi_i;\Phi,J]=
\int\limits_{\phi_f\phi_i}\D\phi
e^{i\left(S_3[\phi,\Phi]+\int_{-T/2}^{T/2}dt\,\trans{\phi}J\right)}
\end{equation}
and involve the modified ``Lagrangian''
\begin{equation}
L_Z[\phi,\Phi,J]
=L_3[\phi,\Phi]+\trans{J}\phi=\frac{1}{2}
\left[
        \trans{\dot{\phi}}m\dot{\phi}+2\trans{\dot{\phi}}\wt{y}
        -
\trans{\phi}\varpi\phi-2\trans{\phi}(\wt{x}-J)
        +\frac{d}{dt}\left(\trans{\phi}\dot{m}\phi\right)
\right].
\end{equation}
We might try to carry out the same completion of the square as was
done in Sec.~\ref{ssec:line}, getting the form \eqref{Lphiz}, where
now
\begin{equation}\tag{\ref{xdefz}$'$}
x=\wt{x}+\dot{z}-J.
\end{equation}
The ODE for $z$ becomes
\begin{equation}\tag{\ref{zDE}$'$}\label{zDEp}
\frac{d}{dt}(\varpi^{-1}\dot{z})+m^{-1}z
=m^{-1}\wt{y}-\frac{d}{dt}[\varpi^{-1}(\wt{x}-J)];
\end{equation}
once again, we would find an expression like
\begin{equation}\tag{\ref{K3X}$'$}\label{K3Xp}
\mc{K}_Z
(\phi_f|\phi_i;\Phi,J]
=\mc{K}_\off(\phi_f|\phi_i;\Phi]
\exp\left\{
        i\trans{\left(\begin{array}{cccc}\phi_f\\ \phi_i\end{array}\right)}
        \mc{X}[\Phi,J]
\right\}
e^{i\psi[\Phi,J]},
\end{equation}
with the expressions \eqref{psidef} and \eqref{Xdef} for
$\psi[\Phi,J]$ and $\mc{X}[\Phi,J]$ in terms of $x$ still holding.
Again, the boundary conditions on \eqref{K3Xp} would allow us to set
$\mc{X}[\Phi,J]=0$, leaving
\begin{equation}\tag{\ref{K3}$'$}
\mc{K}_Z(\phi_f|\phi_i;\Phi,J]
=\mc{K}_\off(\phi_f|\phi_i;\Phi]e^{i\psi[\Phi,J]}.
\end{equation}
However, this form is not convenient, even if we insert
$x=\wt{x}+\dot{z}-J$, since the expression would depend on $J$
not only explicitly, but also implicitly via the solution $z[\Phi,J]$
to \eqref{zDEp}.

\subsubsection{The correct way to complete the square}

        Since we cannot fruitfully complete the square for the
$J$-terms in the way we did for $L_\phi$ in Sec.~\ref{ssec:line}, let
us instead combine $L_\phi$ with the $J$-terms by integrating
by parts until $y=0$, {i.e.},
\begin{mathletters}
\begin{eqnarray}
x&=&\wt{x}+\dot{\wt{y}}
\\
z&=&\wt{y}
\end{eqnarray}
\end{mathletters}
so that
\begin{equation}
L_Z
[\phi,\Phi,J]
=\frac{1}{2}\left[
        \trans{\dot{\phi}}m\dot{\phi}
        -\trans{\phi}\varpi\phi+2\trans{\phi}\wt{J}
        +\frac{d}{dt}\left(\trans{\phi}\dot{m}\phi
                +2\trans{\phi}\wt{y}\right)
\right]
\end{equation}
where
$\wt{J}=J-\wt{x}+\dot{\wt{y}}$.

        First, we define a Green's function $G(t,t')$ (implicitly
dependent upon $\Phi$) satisfying
\begin{equation}
\left[\partial_t m(t)\partial_t+\varpi(t)\right]G(t,t')=\delta(t-t'),
\end{equation}
so that
\begin{equation}
(G\circ J)(t)=\int_{-T/2}^{T/2}dt'\, G(t,t')J(t')
\end{equation}
obeys $\left(\partial_t m\partial_t+\varpi\right)G\circ J=J$.
We can construct this perturbatively, with the lowest order term being
\begin{equation}
G_0(t,t')=\frac{\sin\Oo\abs{t-t'}}{2\Oo}.
\end{equation}
Then we can complete the square to obtain
\begin{eqnarray}
\mc{K}_Z(\phi_f|\phi_i;\Phi,J]&=&\mc{K}_\off\left(\phi_f-(G\circ\wt{J})_f
        \right|\left.\phi_i-(G\circ\wt{J})_i\right)
\nonumber\\
&&\times
\exp\left(i\left\{
        -S_\off[G\circ\wt{J},\Phi]
        +\left.\trans{\phi}[\partial_t(mG\circ\wt{J})+\wt{y}]
         \right\rvert_{-T/2}^{T/2}
\right\}\right).
\end{eqnarray}
The generating functional can thus be expanded, using the
form of $\rho_\phi$ from \eqref{rhodef}, and making the
transformation ({cf.}\ \eqref{bardef})
\begin{equation}
\left(\begin{array}{cccc}
        \phi_f \\ \phi'_f \\ \phi_i \\ \phi'_i
\end{array}\right)
=\left(\begin{array}{cccc} 
1 & 1 & 0 & 0 \\-1 & 1 & 0 & 0 \\ 0 & 0 & 1 & 1 \\ 0 & 0 & 1 &-1
\end{array}\right)
\left(\begin{array}{cccc}
         \Delta\phi_f/2 \\ \overline{\phi_f} \\ \overline{\phi_i} \\ \Delta\phi_i/2
\end{array}\right),
\end{equation}
as
\begin{eqnarray}
Z[J,J',\Phi,\Phi']
\propto\int\frac{d\overline{\phi_i}d\Delta\phi_id\phi_f}
{\sqrt{\det(2\pi\Crune[\Phi])\det(2\pi\Crune[\Phi'])}}
&&\exp\left\{
        -\frac{1}{2}
\trans{
  \left(\begin{array}{cccc}
        \phi_f \\ \overline{\phi_i} \\ \Delta\phi_i/2
  \end{array}\right)
}
\mc{M}
\left(\begin{array}{cccc}
        \phi_f \\ \overline{\phi_i} \\ \Delta\phi_i/2
\end{array}\right)
\right.
\nonumber\\
+\trans{
  \left(\begin{array}{cccc}
        \phi_f \\ \overline{\phi_i} \\ \Delta\phi_i/2
  \end{array}\right)
}
&&\left.
        \left[
                (\mc{M}-\mc{P})\mc{U}[\wt{J}]
                +i\mc{W}[\wt{J}]
                +i\left(\begin{array}{cccc}
                        \Delta\wt{y}_f \\
                        -\Delta\wt{y}_i \\
                        -2\overline{\wt{y}_i}
                \end{array}\right)
        \right]
+i\mc{Q}[\wt{J},\wt{J}']
\right\},\label{Zphi}
\end{eqnarray}
where we have defined the matrix
\begin{equation}
\mc{P}=
\left(\begin{array}{cccc}
0 & 0 & 0 \\
0 & \Oo^2\mc{V}(\Oo) & 0 \\
0 & 0 & 4\mc{V}^{-1}(\Oo)
\end{array}\right)
\end{equation}
so that $\mc{M}$,  defined in \eqref{Mdef}, can be written
\begin{equation}
\mc{M}=\mc{P}-i
\left(\begin{array}{cccc}
\mc{B}_{-} & -\mc{C}_{-} & -\mc{C}_{+} \\
-\trans{\mc{C}_{-}} & \mc{A}_{-} & \mc{A}_{+} \\
-\trans{\mc{C}_{+}} & \mc{A}_{+} & \mc{A}_{-}
\end{array}\right),
\end{equation}
as well as
\begin{mathletters}
\begin{eqnarray}
\mc{U}[\wt{J}]&=&\left(\begin{array}{cccc}
        \overline{(G\circ\wt{J})_f} \\
        \overline{(G\circ\wt{J})_i} \\
        \Delta(G\circ\wt{J})_i/2
\end{array}\right)\label{Udef}
\\
\mc{W}[\wt{J}]&=&\left(\begin{array}{cccc}
        -\mc{B}_{+}\Delta(G\circ\wt{J})_f/2
                +\Delta\left[\partial_t(mG\circ\wt{J})\right]_f \\
        \trans{\mc{C}_{+}}\Delta(G\circ\wt{J})_f/2
                -\Delta\left[\partial_t(mG\circ\wt{J})\right]_i \\
        \trans{\mc{C}_{-}}\Delta(G\circ\wt{J})_f/2
                 -2\overline{\left[\partial_t(mG\circ\wt{J})\right]_i} 
\end{array}\right)\label{Wdef}
\\
\mc{Q}[\wt{J},\wt{J}']
&=&-S_\off[G\circ\wt{J},\Phi]+S_\off[G'\circ\wt{J}',\Phi'].
\nonumber\\
&&+\frac{1}{2}
\trans{
  \left(\begin{array}{cccc}
        \Delta(G\circ\wt{J})_f/2 \\
        \overline{(G\circ\wt{J})_f} \\
        \overline{(G\circ\wt{J})_i} \\
        \Delta(G\circ\wt{J})_i/2
  \end{array}\right)
}
\left(\begin{array}{cccc}
\mc{B}_{-} & \mc{B}_{+} & -\mc{C}_{+} & -\mc{C}_{-} \\
\mc{B}_{+} & \mc{B}_{-} & -\mc{C}_{-} & -\mc{C}_{+} \\
-\trans{\mc{C}_{+}} & -\trans{\mc{C}_{-}} & \mc{A}_{-} & \mc{A}_{+} \\
-\trans{\mc{C}_{-}} & -\trans{\mc{C}_{+}} & \mc{A}_{+} & \mc{A}_{-}
\end{array}\right)
\left(\begin{array}{cccc}
        \Delta(G\circ\wt{J})_f/2 \\
        \overline{(G\circ\wt{J})_f} \\
        \overline{(G\circ\wt{J})_i} \\
        \Delta(G\circ\wt{J})_i/2
\end{array}\right)
\end{eqnarray}
\end{mathletters}
Completing the square in \eqref{Zphi} and integrating gives
\begin{eqnarray}
Z[J,J',\Phi,\Phi']
=e^{iW_\off[\Phi,\Phi']}
&&\exp\left\{
        \frac{1}{2}\left[
                \trans{\mc{U}[\wt{J}]}(\mc{M}-\mc{P})
                +i\trans{\mc{W}[\wt{J}]}
                +i\trans{\left(\begin{array}{cccc}
                        \Delta\wt{y}_f \\
                        -\Delta\wt{y}_i \\
                        -2\overline{\wt{y}_i} 
                \end{array}\right)}
        \right]
\right.\nonumber\\
&&\left.
        \times  \mc{M}^{-1}
        \left[
                (\mc{M}-\mc{P})\mc{U}[\wt{J}]
                +i\mc{W}[\wt{J}]
                +i\left(\begin{array}{cccc}
                        \Delta\wt{y}_f \\
                        -\Delta\wt{y}_i \\
                        -2\overline{\wt{y}_i} 
                \end{array}\right)
        \right]
+i\mc{Q}[\wt{J},\wt{J}']
\right\}\label{Zcs}
\end{eqnarray}

        Expanding \eqref{degen} in a perturbation series, we see that
terms beyond the zeroth have at least one factor of $\ell$, from the
$\ell S_\fff$.  Again, the only way that a perturbative expression
could affect the non-perturbative result $e^{iW}\ll 1$ is if some of
the terms have a $\ell/\beta$ behavior.  Thus, we should look for the
terms in the exponential of \eqref{Zcs} which are larger than
$\mc{O}(1)$ to see if any $\mc{O}(\beta^{-1})$ terms can produce
significant contributions.  The only object which can be larger than
$\mc{O}(1)$ is $\mc{M}^{-1}$ [the matrices $\mc{M}$ and $\mc{P}$
individually have $\mc{V}^{-1}$ eigenvalues, but the combination
$\mc{M}-\mc{P}$ is $\mc{O}(1)$].  Since the smallest eigenvalue of
$\mc{M}$ is $\mc{O}(\mc{V})+\mc{O}(\ell)$, $\ell\mc{M}^{-1}$ will also
be no larger than $\mc{O}(1)$.  And since the terms $\wt{y}$ and
$\wt{J}-J=-\wt{x}+\dot{\wt{y}}$ coming from $S_\phi$ are
$\mc{O}(\ell)$, this means that 
\begin{equation}
Z[J,J',\Phi,\Phi']
=e^{iW_\off[\Phi,\Phi']}
\exp\left(
        \frac{1}{2}\left\{
                \trans{\mc{U}[J]}(\mc{M}-\mc{P})
                +i\trans{\mc{W}[J]}
        \right\}
        \mc{M}^{-1}
        \left\{
                (\mc{M}-\mc{P})\mc{U}[J]
                +i\mc{W}[J]
        \right\}
        +\mc{O}(1)
\right).
\end{equation}
Now,
\begin{equation}
\left[\trans{\mc{U}}(\mc{M}-\mc{P})+i\trans{\mc{W}}\right]
\mc{M}^{-1}\left[(\mc{M}-\mc{P})\mc{U}+i\mc{W}\right]
=-\trans{\mc{U}}\mc{P}\mc{U}
+(\trans{\mc{U}}\mc{P}-i\trans{\mc{W}})\mc{M}^{-1}(\mc{P}\mc{U}-i\mc{W})
+\mc{O}(1);
\end{equation}
if we use \eqref{Mtilde} to write $\mc{M}^{-1}$ in terms of
$\wt{\mc{M}}^{-1}$
and manipulate $\mc{M}^{-1}\mc{P}$ using
$4\mc{V}^{-1}(\Oo)=\alpha+i\mc{A}_{-}$,
we have
\begin{eqnarray}
-\trans{\mc{U}}&&\mc{P}\mc{U}
+(\trans{\mc{U}}\mc{P}-i\trans{\mc{W}})\mc{M}^{-1}(\mc{P}\mc{U}-i\mc{W})
\nonumber\\
=&&-\trans{\mc{U}}\mc{P}\mc{U}
+\left[\trans{\mc{U}}
\left(\begin{array}{ccc}
0 & 0 & 0 \\
0 & \Oo^2\mc{V}(\Oo) & 0 \\
-i(1+i\mc{A}_{-}\alpha^{-1})\trans{\mc{C}}_{+}
        & i(1+i\mc{A}_{-}\alpha^{-1})\mc{A}_{+} & 4\mc{V}^{-1}(\Oo)
\end{array}\right)
\right.
\left.
-i\trans{\mc{W}}
\left(\begin{array}{ccc}
1 & 0 & 0 \\
0 & 1 & 0 \\
-i\alpha^{-1}\trans{\mc{C}_{+}} & i\alpha^{-1}\mc{A}_{+} & 1
\end{array}\right)
\right]
\nonumber\\
&&\times\wt{M}^{-1}
\left[
\left(\begin{array}{ccc}
0 & 0 & -i\mc{C}_{+}(1+i\alpha^{-1}\mc{A}_{-}) \\
0 & \Oo^2\mc{V}(\Oo) & i\mc{A}_{+}(1+i\alpha^{-1}\mc{A}_{-}) \\
0 & 0 & 4\mc{V}^{-1}(\Oo)
\end{array}\right)
\mc{U}-i
\left(\begin{array}{ccc}
1 & 0 & -i\mc{C}_{+}\alpha^{-1} \\
0 & 1 & i\mc{A}_{+}\alpha^{-1} \\
0 & 0 & 1
\end{array}\right)
\mc{W}
\right].
\label{UPUbig}\end{eqnarray}
Because the matrices $\mc{P}$ and
$(\aleph_0-i\ell\aleph_1[\Delta\Phi])^{-1}\oplus\alpha^{-1}$ are in
block diagonal form, we split up the expression \eqref{UPUbig} into
parts corresponding to each block.  That corresponding to the lower
block is
\begin{equation}\label{blockb}
-\trans{\mc{U}}_3 4\mc{V}^{-1}(\Oo)\mc{U}_3
+\left[
        \trans{\mc{U}}_3 4\mc{V}^{-1}(\Oo)-i\trans{\mc{W}}_3
\right]
\alpha^{-1}
\left[
        4\mc{V}^{-1}(\Oo)\mc{U}_3-i\mc{W}_3
\right],
\end{equation}
where $\mc{U}_3$ is the bottom third of $\mc{U}$,
\begin{equation}
\mc{U}_3=\trans{\left(\begin{array}{cccc}0 \\ 0 \\ 1\end{array}\right)}\mc{U}.
\end{equation}
Using $4\mc{V}^{-1}(\Oo)=\alpha+i\mc{A}_{-}$, the $\mc{V}^{-1}(\Oo)$
pieces of \eqref{blockb} cancel, leaving an expression which is
$\mc{O}(1)$.
        This leaves us with the piece from the upper block,
making the exponential in \eqref{Zcs}
\begin{eqnarray}
\frac{1}{2}&&\left[\trans{\mc{U}}
\left(\begin{array}{cc}
0 & 0 \\
0 & \Oo^2\mc{V}(\Oo) \\
-i\trans{\mc{C}}_{+} & i\mc{A}_{+}
\end{array}\right)
-i\trans{\mc{W}}
\left(\begin{array}{cc}
1 & 0 \\
0 & 1 \\
-i\alpha^{-1}\trans{\mc{C}_{+}} & i\alpha^{-1}\mc{A}_{+}
\end{array}\right)
\right]
(\aleph_0-i\ell\aleph_1[\Delta\Phi])^{-1}
\nonumber\\
&&\phantom{\left[\trans{\mc{U}}
   \left(\begin{array}{cc}
   0 & 0 \\
   0 & \Oo^2\mc{V}(\Oo) \\
   -i\trans{\mc{C}}_{+} & i\mc{A}_{+}
   \end{array}\right)\right]
}
\times
\left[
\left(\begin{array}{ccc}
0 & 0 & -i\mc{C}_{+} \\
0 & \Oo^2\mc{V}(\Oo) & i\mc{A}_{+}
\end{array}\right)
\mc{U}-i
\left(\begin{array}{ccc}
1 & 0 & -i\mc{C}_{+}\alpha^{-1} \\
0 & 1 & i\mc{A}_{+}\alpha^{-1}
\end{array}\right)
\mc{W}
\right]
+\mc{O}(1);
\end{eqnarray}
inserting \eqref{Udef} and \eqref{Wdef}
and discarding $\mc{O}(1)$ terms \{including
$(\aleph_0-i\ell\aleph_1[\Delta\Phi])^{-1}\mc{V}(\Oo)$\},
we end up with
\begin{eqnarray}
Z[J,J',\Phi,\Phi']
=&&e^{iW_\off[\Phi,\Phi']}
\exp\left[-\frac{1}{2}
\trans{\left(\begin{array}{cccc}
        -C_0 G_0\circ(\Delta J)_i+B_0 G_0\circ(\Delta J)_f
        -\dot{G}_0\circ(\Delta J)_f \\
        A_0 G_0\circ(\Delta J)_i-C_0 G_0\circ(\Delta J)_f
        +\dot{G}_0\circ(\Delta J)_i
\end{array}\right)
}\right.
\nonumber\\
&&\phantom{e^{iW}}
\left.\times
(\aleph_0-i\ell\aleph_1[\Delta\Phi])^{-1}
\left(\begin{array}{cccc}
-C_0 G_0\circ(\Delta J)_i+B_0 G_0\circ(\Delta J)_f
        -\dot{G}_0\circ(\Delta J)_f \\
A_0 G_0\circ(\Delta J)_i-C_0 G_0\circ(\Delta J)_f
        +\dot{G}_0\circ(\Delta J)_i
\end{array}\right)
+\mc{O}(1)
\right].\label{Zlead}
\end{eqnarray}

The fact that the leading term in the exponential in \eqref{Zlead}
depends only upon $\Delta J=J-J'$ is crucial, because 
of the operator
\begin{equation}
\ell S_\fff\left[\frac{1}{i}\frac{\D}{\D J}\right]
-\ell S_\fff\left[-\frac{1}{i}\frac{\D}{\D J'}\right]
\end{equation}
in \eqref{degen}, which annihilates any functional independent of
$\phi$ and depending on $J^{(\prime)}$ only in the combination $J-J'$.
If \emph{all} of the terms in the exponential in $Z$ were functions of
$\Delta J$ alone, that would mean that $e^{iW}=e^{iW_3}$; however,
there are $\mc{O}(1)$ terms in the exponential which depend on
$\overline{J}$.  The situation can be written as
\begin{equation}
Z[J,J',\Phi,\Phi']
=e^{iW_\off[\Phi,\Phi']}
\exp\left(\frac{1}{2}\Delta J\circ\mc{F}_{-1}\circ\Delta J
        + \frac{1}{2}J^{(\prime)}\circ\mc{F}_0\circ J^{(\prime)}
        + J^{(\prime)}\circ\mc{G}_0
\right),
\end{equation}
where $\frac{1}{2}\Delta J\circ\mc{F}_{-1}\circ\Delta J$ is the
argument of the exponential in \eqref{Zlead}, and
$\frac{1}{2}J^{(\prime)}\circ\mc{F}_0\circ J^{(\prime)}$ and
$J^{(\prime)}\circ\mc{G}_0$ are the quadratic and linear terms of
$\mc{O}(1)$.  Thinking in terms of a diagrammatic expansion, this
means that there are three kinds of ``propagators'' in $Z$:
\begin{mathletters}
\begin{eqnarray}
\psfig{figure=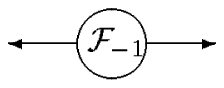}
\label{graph1}
\\
\psfig{figure=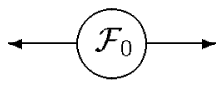}
\label{graph2}
\\
\psfig{figure=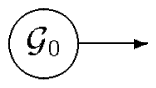}
\label{graph3}
\end{eqnarray}
\end{mathletters}
(note that the last is not truly a propagator, since
it accepts only one ``input'').  These are used to connect the
vertices, which all have the form
\begin{equation}\label{graph4}
\psfig{figure=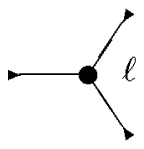}
\end{equation}
A term in the series which has more $\ell$ vertices than
$\mc{F}_{-1}$ propagators will be perturbatively small, one with the
same number will be $\mc{O}(1)$, and one with more $\mc{F}_{-1}$
propagators than $\ell$ vertices will be able to disrupt the
perturbative analysis and have an impact upon $e^{iW}$.  We can make a
list of the objects in the theory by their order in perturbation
theory and number of legs (with the legs on propagators counted
negative so that a closed diagram has zero net legs):
\begin{center}
\begin{tabular}{lcc}
Graph          & Order & Legs \\
\eqref{graph1} &  -1  &  -2\\
\eqref{graph2} &   0  &  -2\\
\eqref{graph3} &   0  &  -1\\
\eqref{graph4} &   1  &  3\\
\end{tabular}
\end{center}
Since the vertex \eqref{graph4} has three legs and the propagator
\eqref{graph1} has minus two, we'd expect divergent graphs starting
with
\begin{equation}\label{graph5}
\psfig{figure=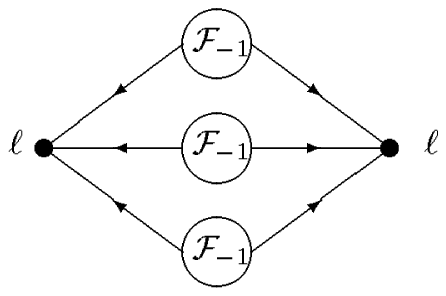}
\end{equation}
However, in this case we have just the situation described above: all
of the propagators depend only on $\Delta J$, so the graph vanishes.
This sort of identity places the restriction that at least one leg of
a vertex must be coupled to an $\mc{F}_0$ or $\mc{G}_0$ propagator.
This means that we must abandon \eqref{graph4} by itself and use as
our primitive vertices
\begin{mathletters}
\begin{eqnarray}
\psfig{figure=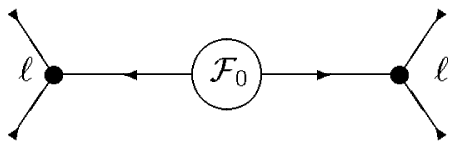}
\label{graph6}
\\
\psfig{figure=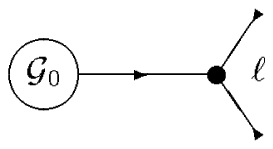}
\label{graph7}
\end{eqnarray}
\end{mathletters}
which makes the pieces out of which non-vanishing graphs can be
constructed
\begin{center}
\begin{tabular}{lcc}
Graph          & Order & Legs \\
\eqref{graph1} &  -1  &  -2\\
\eqref{graph2} &   0  &  -2\\
\eqref{graph3} &   0  &  -1\\
\eqref{graph6} &   2  &  4\\
\eqref{graph7} &   1  &  2 
\end{tabular}
\end{center}
Now the most divergent graph which can be constructed with zero net
legs is $\mc{O}(1)$.

        This means that, perturbatively, the influence functional
is
\begin{equation}\tag{\ref{eiWdiff}}
e^{iW[\Phi,\Phi']}=\mc{O}(1)\times e^{iW_3[\Phi,\Phi']},
\end{equation}
so, perturbatively at least,
\begin{equation}\tag{\ref{eiWres}}
\abs{e^{iW[\Phi,\Phi']}}
\lesssim\left\{
        1+E^2[\Delta\Phi]
\right\}^{-1/4}.
\end{equation}

\end{document}